\documentclass[11pt]{article}
\usepackage{amsmath}
\usepackage{amssymb}
\usepackage{amsfonts}
\usepackage{amscd}
\usepackage{accents}
\usepackage{bbm} % BlackBoeard letters
\usepackage{pst-all}
\usepackage{epsfig}
\date{\today}
\newcommand{\bmat}{\left(\begin{array}}
\newcommand{\emat}{\end{array}\right)}
\newcommand{\be}{\begin{equation}}
\newcommand{\ee}{\end{equation}}
\newcommand{\ba}{\begin{eqnarray}}
\newcommand{\ea}{\end{eqnarray}}

\def\lsim{\raise0.3ex\hbox{$\;<$\kern-0.75em\raise-1.1ex\hbox{$\sim\;$}}}
\def\gsim{\raise0.3ex\hbox{$\;>$\kern-0.75em\raise-1.1ex\hbox{$\sim\;$}}}

\def\be{\beta}

\begin{document}
\vspace*{-.6in} \thispagestyle{empty}
\begin{flushright}
CERN-PH-TH/2011-106\\
DESY 11-075
\end{flushright}
\baselineskip = 20pt

\vspace{.5in} {\Large
\begin{center}
{\bf Higgs Portal Inflation}

\end{center}}

\vspace{.5in}

\begin{center}
{\bf  Oleg Lebedev$^{~a}$  and  
 Hyun Min Lee$^{~b} $}  \\

\vspace{.5in}

a: \emph{DESY Theory Group, 
Notkestrasse 85, D-22607 Hamburg, Germany
 }
\\ 
b: \emph{CERN, Theory Division, CH-1211 Geneva 23, Switzerland  }
\end{center}

\vspace{.5in}

\begin{abstract}
The Higgs sector of the Standard Model offers a unique opportunity 
to probe the hidden sector. The  Higgs squared  operator 
is the only  dimension two operator which is  Lorentz and gauge 
invariant. It can therefore couple both to scalar curvature and 
the hidden sector at the dim--4 level. We consider the possibility
that a combination of the Higgs and a singlet from the hidden sector 
plays the role of  inflaton, due to their large couplings to gravity. 
This implies that  the quartic   couplings satisfy certain 
constraints which leads to   distinct   low energy
phenomenology, including Higgs signals at the LHC.
We also address the unitarity issues and show that our analysis
survives the unitarization procedure.
\end{abstract}

\noindent

\newpage

\section{Introduction}

Cosmic inflation \cite{inflation} is  a paradigm 
beyond the Standard Big Bang Cosmology  which addresses the 
flatness, isotropy, homogeneity, horizon and relic problems.
Furthermore, quantum fluctuations during inflation provide a seed for the 
large--scale structure formation. On the other hand, the nature of 
the inflaton remains a mystery.
It has recently been conjectured that the only scalar   of the 
Standard Model (SM),
the Higgs field,  may play its role \cite{Bezrukov:2007ep}, given 
 a large Higgs coupling to scalar curvature.
The Higgs sector is also quite special because it has a direct access to the
``hidden sector'' \cite{Patt:2006fw}, 
whose existence is motivated by various ideas including 
string theory,  dark matter, etc.  Understanding the Higgs couplings would
thus provide us with unique information about the hidden world.

There are two dim-2 operators in the Standard Model that can couple
to the hidden sector at the renormalizable level: $F_{\mu\nu}^Y$ and
$H^\dagger H$. The latter is also Lorentz invariant, so it can in addition
couple to scalar curvature $R$. One can therefore add the following
dim--4 operators to the Standard Model Lagrangian,
\begin{eqnarray}
&& \Delta {\cal L}_1 = c_1~ H^\dagger H \vert S \vert^2 \;, \nonumber\\
&& \Delta {\cal L}_2 = c_2~ H^\dagger H R \;,
\end{eqnarray}
where $S$ is a singlet under the Standard Model and $c_i$ are dimensionless
constants. 
In what follows, we consider the minimal option for the hidden sector: 
we take $S$ to be a real scalar $s$ and impose the symmetry 
$s\leftrightarrow -s$.
The coupling $c_1$ controls the Higgs decay into the hidden sector as well
as the Higgs--singlet mixing, which can be measured at the LHC.
$c_2$ can be responsible for inflation: with $\vert c_2 \vert \gg 1$,
 a large  value of the Higgs field in the early universe leads to 
exponential expansion.

In this work, we consider the possibility that the inflaton is a mixture
of the Higgs with the singlet from the hidden sector. The nature of 
the inflaton depends on the relations among various couplings. For example,
if $c_1$ is positive, stability of the potential requires a mixed 
inflaton. On the other hand, for negative $c_1$ the inflaton can be purely
the Higgs or the singlet field. These considerations leave an imprint 
on the low energy physics, affecting the couplings of the Higgs--like
particles to be studied at the LHC.

We also study the unitarity issues which plague the original
Higgs inflation \cite{unitarity1,unitarity2}. We construct a unitary 
completion \cite{giudicelee}   of the Higgs portal inflation and show that the 
constraints on the couplings survive the unitarization procedure. 

The paper is organized as follows. We first present a general analysis
of the SM extension with a real singlet in the presence of large
couplings to scalar curvature. We study stability of the system during 
inflation and derive the corresponding constraints on the couplings.
Then we study implications for low energy physics. 
We further discuss the differences from   the pure Higgs \cite{Clark:2009dc}  and
 singlet inflation \cite{singletinflation,Okada:2010jd}, and present an example of the
unitary completion of our model.

\section{Higgs--singlet combination as the inflaton}

In this section, we study an extension of the Higgs sector with a real 
scalar $s$ in the presence of large couplings $\xi_{h,s}$ to scalar
 curvature $R$. This system can lead to inflation   
  based on scale invariance of the  Einstein frame  
scalar potential at large field values.
The relevant Jordan frame Lagrangian in the unitary gauge 
$H^T=(0,h/\sqrt{2})$
is
\begin{equation}
{\cal L}/ \sqrt{-g} =  - {1\over 2} M^2_{\rm Pl} R - {1\over 2}   \xi_h h^2 R - {1\over 2}   \xi_s s^2 R    
+ {1\over 2} (\partial_\mu h)^2 + {1\over 2} (\partial_\mu s)^2
- V                
\end{equation}
with  $\xi_{h,s} >0$\footnote{We do not consider negative $\xi_i$ since
in this case the theory is not well defined at large field values.} and 
\begin{equation}
V= {1\over 4} \lambda_h h^4 + {1\over 4}\lambda_{hs} s^2 h^2 + {1\over 4}\lambda_s s^4 + 
{1\over 2} m_h^2 h^2 + {1\over 2} m_s^2 s^2 \;. \label{V}
\end{equation}
The transformation to the Einstein frame, in which the only  coupling
to curvature is $-1/2 M_{\rm Pl}^2 R$,
 is defined by
\begin{equation}
\tilde g_{\mu\nu}=\Omega^2 g_{\mu\nu} ~~,~~
\Omega^2= 1 + { \xi_h h^2 +  \xi_s s^2 \over M_{\rm Pl}^2} \;. \label{Omega}
\end{equation}
Consider now the limit 
\begin{equation}
\xi_h h^2 +  \xi_s s^2 \gg  M_{\rm Pl}^2 
\end{equation}
and set $M_{\rm Pl}$ to 1. In this case, $\Omega^2 \simeq 
\xi_h h^2 +  \xi_s s^2 $.
Then, according to \cite{Salopek:1988qh},
the kinetic terms and the potential in the  Einstein frame take the form
\begin{eqnarray}
{\cal L}_{\rm kin} &=&  
{3\over 4} \biggl(  \partial_\mu \log (\xi_h h^2 +  \xi_s s^2)    \biggr)^2
+{1\over 2} {1\over \xi_h h^2 +  \xi_s s^2} \biggl(  (\partial_\mu h)^2 +
(\partial_\mu s)^2   \biggr) \;, \nonumber\\
U &=& {1\over (\xi_h h^2 +  \xi_s s^2)^2} ~V \;. \label{Lkin}
\end{eqnarray}
Introduce new variables
\begin{eqnarray}
&& \chi =  \sqrt{3\over 2}  ~ \log (\xi_h h^2 +  \xi_s s^2) \;, \nonumber\\
&& \tau = {h\over s} \;. 
\end{eqnarray}
In terms of these variables, the kinetic terms read
\begin{eqnarray}
{\cal L}_{\rm kin} &=& {1\over 2} \biggl( 1+ {1\over 6} {\tau^2 +1\over
\xi_h \tau^2 +\xi_s} \biggr)~ (\partial_\mu \chi)^2 + 
{1\over \sqrt{6}} ~{(\xi_s-\xi_h) \tau \over (\xi_h \tau^2 +\xi_s)^2} 
(\partial_\mu \chi) (\partial^\mu \tau) \nonumber\\
&+&
{1\over 2}{ \xi_h^2 \tau^2 +\xi_s^2 \over (\xi_h \tau^2 +\xi_s)^3 }(\partial_\mu \tau)^2
\;. 
\end{eqnarray}
We are interested in the case of large non-minimal couplings,
 $ \xi \equiv \xi_h  +\xi_s \gg 1$. Since the $(\partial_\mu \tau)^2$ 
term scales like $1/\xi$ and so does the mixing term $(\partial_\mu \chi) (\partial^\mu \tau) $, in terms of  (approximately)  canonically normalized variables  the mixing is suppressed. Then, to leading order in $1/\xi$, we have 
 \begin{equation}
{\cal L}_{\rm kin}= {1\over 2 } (\partial_\mu \chi)^2 +
{1\over 2} { \xi_h^2 \tau^2 +\xi_s^2 \over (\xi_h \tau^2 +\xi_s)^3 }(\partial_\mu \tau)^2 \;.
\end{equation}
In the  following limiting cases, one can define a particularly simple
canonically normalized variable $\tau'$ :
\begin{eqnarray}
&& \xi_s \gg \xi_h  ~~{\rm or}~~ \tau \rightarrow 0 ~~,~~~~ \tau'= {\tau\over
\sqrt{\xi_s}}  ~, \nonumber\\
&& \xi_h \gg \xi_s  ~~{\rm or}~~ \tau \rightarrow \infty ~~,~~ \tau'= {1\over
\sqrt{\xi_h} \tau}  ~, \nonumber\\
&& \xi_h = \xi_s  ~~,~~~~~~~~~~~~~~~~~~~  \tau'= {1\over{\sqrt{\xi_h}}} \arctan \tau \;.
\end{eqnarray}
The scalar potential at large $\chi$ reads
\begin{equation}
U= {\lambda_h \tau^4 + \lambda_{hs} \tau^2 +\lambda_s  \over
 4 (\xi_h \tau^2 +\xi_s)^2} \;.
\end{equation}
Its minima are classified according to 
\begin{eqnarray}
&& (1)~2 \lambda_h \xi_s - \lambda_{hs} \xi_h >0~,~
   2 \lambda_s \xi_h - \lambda_{hs} \xi_s >0~,~~~~\tau = \sqrt{ 
 2 \lambda_s \xi_h - \lambda_{hs} \xi_s  \over 
  2 \lambda_h \xi_s - \lambda_{hs} \xi_h   }   \;, \nonumber\\
&& (2)~2 \lambda_h \xi_s - \lambda_{hs} \xi_h >0~,~
   2 \lambda_s \xi_h - \lambda_{hs} \xi_s <0~,~~~~\tau=0  \;, \nonumber\\
&& (3)~2 \lambda_h \xi_s - \lambda_{hs} \xi_h <0~,~
   2 \lambda_s \xi_h - \lambda_{hs} \xi_s >0~,~~~~\tau=\infty  \;, \nonumber\\
&& (4)~2 \lambda_h \xi_s - \lambda_{hs} \xi_h <0~,~
   2 \lambda_s \xi_h - \lambda_{hs} \xi_s <0~,~~~~\tau=0,\infty  \;. \label{taumin}
\end{eqnarray}
Note that in the last case there are 2 local minima. We are primarily
interested in the first case, when the inflaton is a combination of the 
Higgs field and the singlet. The corresponding value of the potential
is then
\begin{equation}
U\Bigl\vert_{\rm min~(1) }= {1\over 16} {4 \lambda_s \lambda_h - \lambda_{hs}^2
\over \lambda_s \xi_h^2 + \lambda_h \xi_s^2 - \lambda_{hs} \xi_s \xi_h } \;,
\label{Umin}
\end{equation}
while in cases (2) and (3), it is $\lambda_s /(4\xi_s^2)$ and 
$\lambda_h /(4\xi_h^2)$, respectively. 
The condition $ 4 \lambda_s \lambda_h - \lambda_{hs}^2 >0$ guarantees the 
absence  of very deep minima with negative vacuum energy at 
field values $m_{h,s} \ll h,s $, which make the electroweak vacuum metastable. With this constraint, the vacuum energy  above is positive
(the denominator is positive by the minimization conditions). 

In all of the cases, the $\tau$-field is heavy and can be integrated out.
Indeed, the mass of the canonically normalized $\tau' $ scales as
$1/ \sqrt{\xi}$ in Planck units, while the Hubble rate scales like 
$ \sqrt{U}\vert_{\rm min} \sim  1/ \xi$. Thus 
\begin{equation} 
m_{\tau'}^2 \gg H^2 \;.
\end{equation}

The potential value (\ref{Umin}) plays the role of the quartic coupling
over $\xi^2$
in the single field inflation model of  Bezrukov--Shaposhnikov 
\cite{Bezrukov:2007ep}.
Retaining the subleading $M_{\rm Pl}^2/(\xi_h h^2 +  \xi_s s^2)$ term
in $\Omega^2$, the inflaton potential for option (1) becomes
\begin{equation}
 U(\chi)= {\lambda_{\rm eff}  \over 4\xi_h^2}~
 \Bigl(    1+ {\rm exp}\left( -{2\chi\over \sqrt{6} } \right)          \Bigr)^{-2}
\end{equation}
in Planck units, where 
\begin{equation}
\lambda_{\rm eff} =   {1\over 4} {4 \lambda_s \lambda_h - \lambda_{hs}^2
\over \lambda_s  + \lambda_h x^2 - \lambda_{hs} x }
\end{equation}
and 
\begin{equation}
x = {\xi_s   \over \xi_h} \;.
\end{equation}

The inflationary parameters are  read off from this potential 
\cite{Bezrukov:2007ep}. 
At large $\chi$, the potential is flat and inflation takes place. 
As $\chi$ rolls to smaller values,  the $\epsilon$-parameter
approaches 1 and inflation ends.
In terms of 
\begin{equation}
\tilde h \simeq {1\over \sqrt{\xi_h} } ~\exp \bigl( \chi /\sqrt{6}  \bigr) \;,
\end{equation} 
 the $\epsilon$-parameter is given by
\begin{equation}
\epsilon = {1\over 2} \left( dU/d\chi \over U  \right)^2 \simeq {4 \over 3 \xi^2_h \tilde h^4 } \;.
\end{equation}
This gives $\tilde h_{\rm end} = (4/3)^{1/4}/ \sqrt{\xi_h}$. Then, for a given number of $e$-folds  
 $N$, the initial value of the inflaton is  $\tilde h_{\rm in} \approx \sqrt{4N/(3\xi_h)}$.
Together with the COBE normalization  $U/\epsilon= 0.027^4$ \cite{Lyth:1998xn}, this fixes $\xi_h$ in terms of
$\lambda_{\rm eff}$, 
 \begin{equation} 
\xi_h \simeq \sqrt{{\lambda_{\rm eff}\over 3 }} ~ {N\over 0.027^2} \;.  \label{xi}
\end{equation}
For $\sqrt{\lambda_{\rm eff}} \sim 1$ and $N=60$, the non-minimal gravity coupling $\xi_h$ is about 
50000. The spectral index is predicted to be
\begin{equation}
n \simeq 1- {2\over N} \simeq 0.97 \;,
\end{equation}
while the tensor to scalar perturbation ratio is $r\simeq 12/N^2 \simeq 0.0033$.
These are robust (tree--level)  predictions of our framework to be tested in the future.\footnote{In multi--field variants of this scenario, large non--Gaussianity can also
be generated \cite{Gong:2011cd}.}
  They are independent of the nature of the inflaton and 
result from the shape of the potential, which in turn follows from  a large 
coupling to scalar curvature.

\subsection{Parameter space analysis}

In this subsection, we analyze the parameter space consistent with the inflaton being
a mixture of the Higgs and singlet fields.
The relavant inflation parameters are evaluated at a high energy scale $\mu$. A 
particular choice of $\mu$ advocated in  \cite{Bezrukov:2008ej}   is to take $\mu\sim m_t(\chi)$
which minimizes the effect of logarithms in the Coleman-Weinberg potential. In
this case, $\mu\sim M_{\rm Pl}/ \sqrt{\xi}$ for large $\chi$.
However, as we discuss in Sec.~\ref{unitaritysection},
the theory is only well defined up to the scale $M_{\rm Pl}/ \xi$ at which 
unitarity violation appears.
We  thus expect new physics to set in  at the 
unitarity scale $\mu_U\sim M_{\rm Pl}/ \xi$ and  take $\mu_U$ as the scale at which
the input parameters are specified. We will assume that the new physics does not
significantly affect the tree level relations of the previous section
(see an example in Sec.~\ref{unitaritysection}), yet it is likely to 
affect the running of the relevant  parameters above $\mu_U$. 
For successful 
Higgs--singlet inflation, we impose at $\mu_U$:
\begin{eqnarray}
&& 2 \lambda_h x - \lambda_{hs}  >0 \;, \nonumber\\
&& 2 \lambda_s {1\over x} - \lambda_{hs}  >0 \;, \nonumber\\
&&  4 \lambda_s \lambda_h - \lambda_{hs}^2 >0 \;. \label{constraints1}
\end{eqnarray}
The third inequality provides an independent constraint for $\lambda_{hs}<0$, while 
for positive $\lambda_{hs}$ it follows from the first two. In addition we 
require perturbativity and stability at $\mu_U$:
\begin{eqnarray}
&& \vert \lambda_i \vert < 1   \;, \nonumber\\
&& \lambda_{h,s} > 0 \;. \label{constraints2}
\end{eqnarray}
Our (judicial) definition of the perturbative couplings is motivated by
perturbativity at the Planck scale. We note that above $ M_{\rm Pl}/ \xi$,
the running of $\lambda_i$ slows down due to the suppression of the inflaton
self-coupling or, equivalently, suppression of its propagator in the Jordan frame
(see, e.g. \cite{singletinflation}). Therefore, our procedure is expected to take  into account the bulk
of  radiative corrections. 
Finally, given uncertainties from new physics above $\mu_U$, 
the running of the parameters, e.g. the spectral index 
\cite{Bezrukov:2008ej,DeSimone:2008ei,Barvinsky:2008ia,Bezrukov:2009db},   
during inflation cannot be reliably calculated
in our framework and we therefore omit  it.

The renormalization group (RG) equations governing the evolution
of couplings below $\mu_U$
are given by \cite{singletinflation}:
\begin{eqnarray}
16\pi^2 {d \lambda_h \over dt}&=& 24 \lambda_h^2 -6 y_t^4 +{3\over 8} \Bigl( 
2 g^4 + (g^2 + g^{\prime 2})^2 \Bigr) \nonumber\\
&+& (-9 g^2 -3 g^{\prime 2}+12 y_t^2) \lambda_h + {1\over 2} \lambda_{hs}^2 \;, \nonumber\\
16\pi^2 {d \lambda_{hs} \over dt} &=& 4 \lambda_{hs}^2 + 12 \lambda_h \lambda_{hs}
-{3\over 2} (3 g^2 + g^{\prime 2}) \lambda_{hs}  \nonumber\\
&+& 6 y_t^2  \lambda_{hs} +
6 \lambda_s  \lambda_{hs} \;, \nonumber\\
 16\pi^2 {d \lambda_{s} \over dt} &=& 2 \lambda_{hs}^2 + 18 \lambda_s^2 \;,
\end{eqnarray}
where $t = \ln (\mu/ m_t)$. The RG equations for the gauge and the top Yukawa couplings
can be found in \cite{Bezrukov:2008ej}. The low energy input values for these couplings are 
$g(m_t)=0.64 , g'(m_t)=0.35 , g_3(m_t)=1.16$, while for the top Yukawa coupling we use
its running value at $m_t$, $y_t (m_t)=0.93$ \cite{Langenfeld:2010aj}.
 For a given set of the low energy couplings 
at $t=0$, we use the above RG equations to run them up to $t \approx 26$ and impose
the constraints (\ref{constraints1}) and  (\ref{constraints2}).

 \begin{figure}[ht]
\epsfig{figure=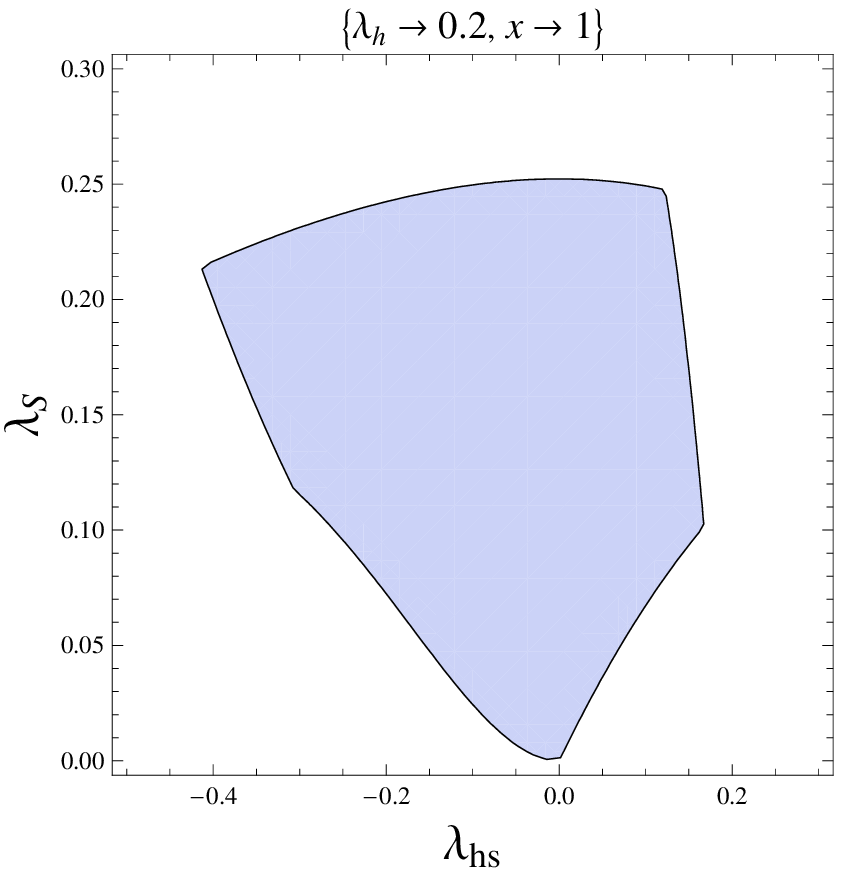,height=5cm,width=5cm,angle=0}
\hspace{0.5cm}\epsfig{figure=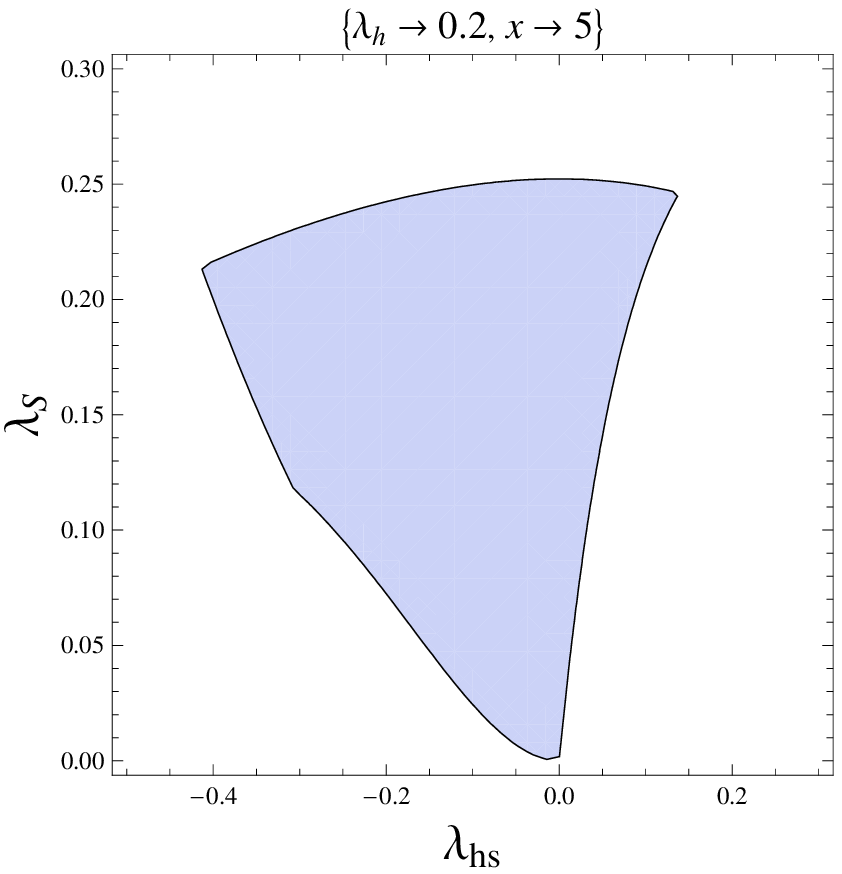,height=5cm,width=5cm,angle=0}
\hspace{0.5cm}\epsfig{figure=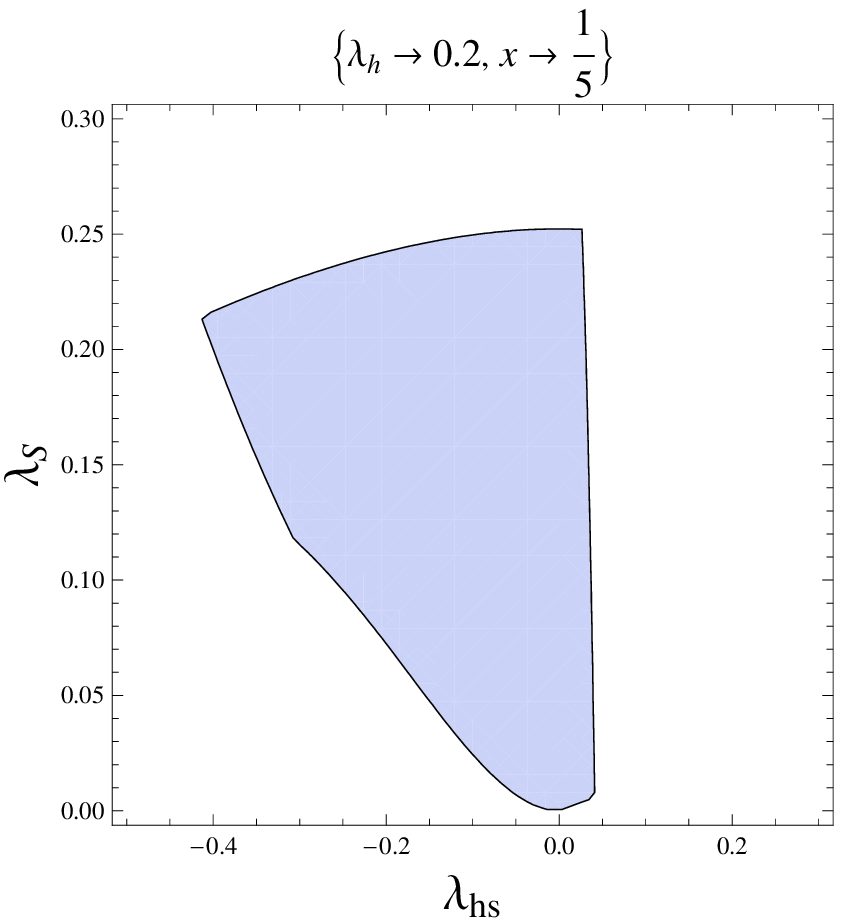,height=5cm,width=5cm,angle=0}
\vspace{0.1cm}\epsfig{figure=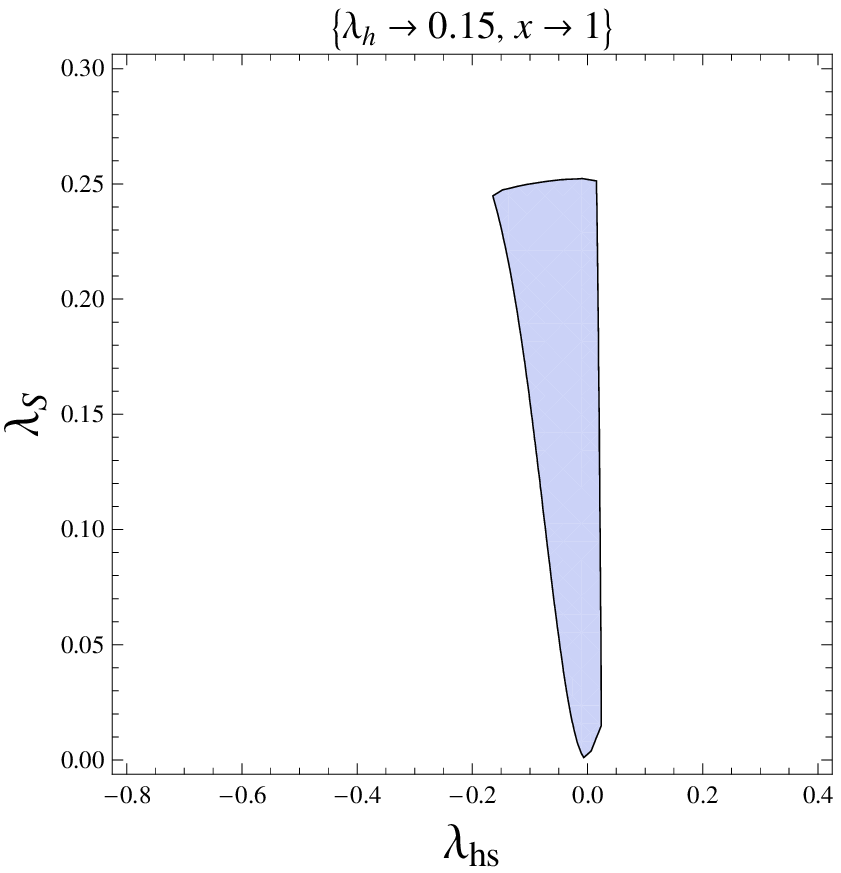,height=5cm,width=5cm,angle=0}
\hspace{0.5cm}\epsfig{figure=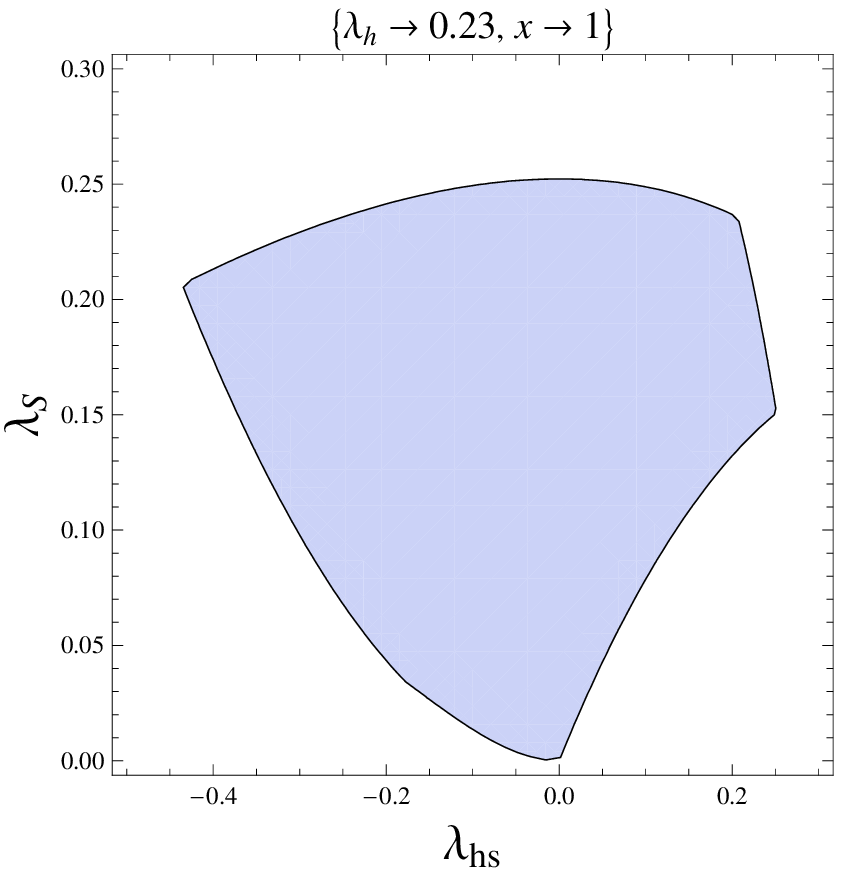,height=5cm,width=5cm,angle=0}
\hspace{0.5cm}\epsfig{figure=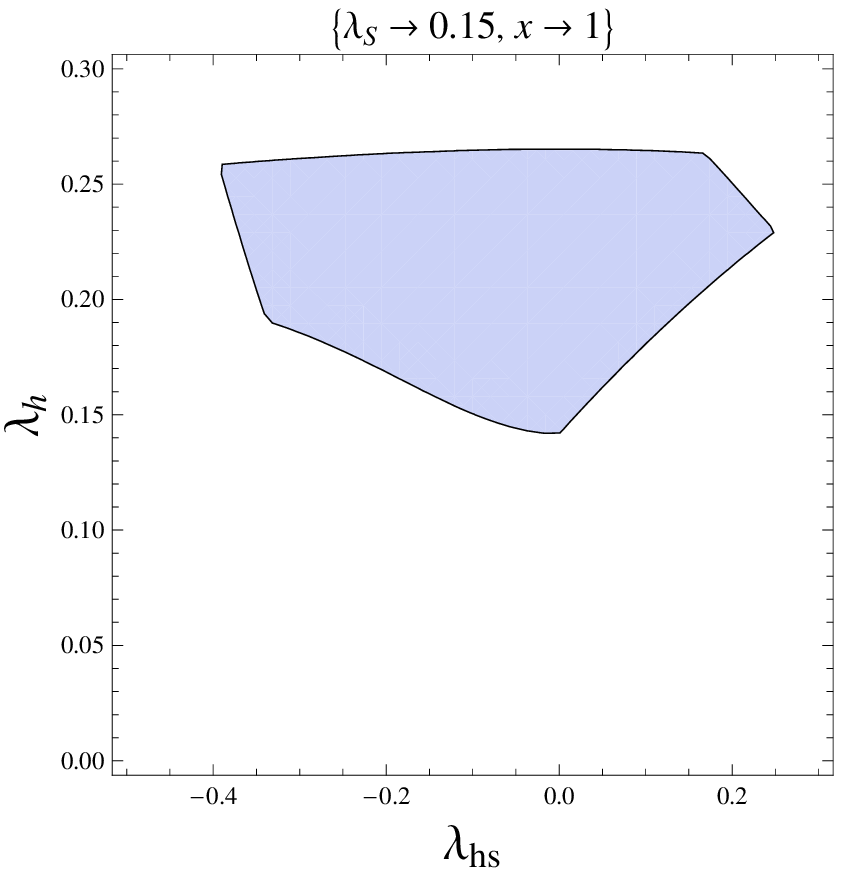,height=5cm,width=5cm,angle=0}
\medskip
\caption{Parameter space consistent with the mixed Higgs--singlet
inflaton. $\lambda_i$ are given at the scale $m_t$, while $x$ is a high energy
input. }
\label{fig1}
\end{figure}

In addition, we impose the low energy constraint at $m_t$ 
\begin{equation}
4 \lambda_s \lambda_h - \lambda_{hs}^2 >0          \label{constraints3}
\end{equation}
for $\lambda_{hs}<0$. This ensures that there are no deep
minima at some intermediate scale $s,h \gg m_s, m_h$ which can make
the electroweak vacuum short-lived.
It is a complementary constraint and (\ref{constraints1}) does not 
guarantee that it is satisfied. We find that for $\lambda_{hs}<0$
the combination $\lambda_s \lambda_h$ can increase with energy  
faster than $\lambda_{hs}^2$ such that parameter space allowed by 
 (\ref{constraints1}) may violate (\ref{constraints3}).

Our results  are presented in Fig.~\ref{fig1}.
In the \{$\lambda_{hs},\lambda_s$\} plane, the parameter space at 
$\lambda_{hs}>0$ is most strongly constrained by 
$ 2 \lambda_s {1\over x} - \lambda_{hs}  >0 $ and, for larger $\lambda_s$, by
$  2 \lambda_h x - \lambda_{hs}  >0  $. In the latter case, $\lambda_s$
contributes significantly to the running of $\lambda_{hs}$, but not to that of
$\lambda_{h}$, which eliminates parameter space to the right  of some critical
value  $\lambda_{hs}$. For negative $\lambda_{hs}$, the main constraint is 
$  4 \lambda_s \lambda_h - \lambda_{hs}^2 >0$ (both at $\mu_U$ and $m_t$)     as well as  perturbativity
 which cuts off large values of $\lambda_s$ and   $\vert \lambda_{hs}\vert$.
At $x\gg 1$ or $x\ll 1$, it becomes more difficult to satisfy either 
 $ 2 \lambda_s {1\over x} - \lambda_{hs}  >0 $ or $  2 \lambda_h x - \lambda_{hs}  >0  $,
so only  small positive values of  $ \lambda_{hs} $ are allowed. On the
other hand, negative $ \lambda_{hs} $ are not affected by $x$.
Decreasing $\lambda_h$ eliminates most of the parameter space and leaves a 
strip around $\lambda_{hs}=0$. The negative top quark   contribution 
to the $\beta$--function of   $\lambda_h$ makes it run slower, reducing 
$\lambda_h (\mu_U)$ and    making it more difficult to
satisfy the constraints at $\mu_U$. 
Naturally, at larger $\lambda_h$, parameter space opens 
up. The range of allowed $\lambda_h$ is similar to that of the Standard Model subject
to the perturbativity and stability requirements, i.e. roughly
$0.14 < \lambda_h < 0.25 $.

Note that the value of $\xi_h$ is not important for our analysis.
Given $\lambda_{\rm eff}$, it is fixed at the scale $\mu_U$ by Eq.(\ref{xi}).
Since we are not interested in its value at low energies, its running
is not relevant for us.

\section{Phenomenological implications}

There are two phenomenologically acceptable possibilities
for the vacuum of our theory:
(a) $\langle h \rangle \not= 0, \langle s \rangle \not= 0$ and 
(b) $\langle h \rangle \not= 0, \langle s \rangle = 0$.
They lead to different phenomenological implications. 

\subsection{ $\langle h \rangle \not= 0, \langle s \rangle \not= 0$ }

Denoting  $\langle h \rangle =v, \langle s \rangle =u$,
extremization of the low energy scalar potential  (\ref{V})   requires
\begin{eqnarray}
&& v^2 = 2~ {   \lambda_{hs} m_s^2 - 2 \lambda_s m_h^2 \over
4 \lambda_s \lambda_h -   \lambda_{hs}^2  } \;, \nonumber\\
&& u^2 = 2~ {   \lambda_{hs} m_h^2 - 2 \lambda_h m_s^2 \over
4 \lambda_s \lambda_h -   \lambda_{hs}^2  } \;.
\end{eqnarray}
The diagonal matrix elements of the Hessian at this point are
$2 \lambda_s u^2$ and  $2 \lambda_h v^2$, while its determinant is
$ (4 \lambda_s \lambda_h -   \lambda_{hs}^2)v^2 u^2  $. Then,
the extremum is a local minimum  if 
\begin{eqnarray}
&& \lambda_{hs} m_h^2 - 2 \lambda_h m_s^2 > 0 \;,\nonumber\\
&& \lambda_{hs} m_s^2 - 2 \lambda_s m_h^2 > 0 \;,\nonumber\\
&& 4 \lambda_s \lambda_h -   \lambda_{hs}^2 >0 \;. \label{ewbreaking}
\end{eqnarray}
In this case, the mass squared eigenvalues are
\begin{equation}
m_{1,2}^2= \lambda_h v^2 + \lambda_s u^2 \mp 
\sqrt{(\lambda_s u^2 - \lambda_h v^2)^2 + \lambda_{hs}^2 u^2 v^2 }
\label{eigenvalues}
\end{equation}
with the mixing angle $\theta$ given by
\begin{equation}
\tan 2 \theta = {\lambda_{hs} u v  \over \lambda_h v^2 - \lambda_s u^2} \;.
\label{tan}
\end{equation}
Here the mixing angle is defined by 
\begin{equation}
O^T~ M^2 ~ O = {\rm diag}(m_1^2, m_2^2) ~~,~~ O=\left( 
\begin{matrix} 
\cos\theta & \sin\theta \\
-\sin\theta & \cos\theta  
\end{matrix} 
\right) \;,
\end{equation}
where $M^2$ is a 2$\times$2 mass squared matrix.
The range of $\theta$ is related to the ordering of the eigenvalues through
$ {\rm sign} (m_1^2-m_2^2 ) = {\rm sign} (\lambda_s u^2 - \lambda_h v^2)~
 {\rm sign} (\cos 2 \theta) $ and we take $m_1$ to be the smaller eigenvalue.
The mass eigenstates are
\begin{eqnarray}
 H_1 &=&  s \cos\theta  - h \sin\theta  \;, \nonumber\\
 H_2 &=&  s \sin\theta  + h \cos\theta  \;. 
\end{eqnarray}
Note that the lighter   mass eigenstate  $H_1$  is 
``Higgs--like'' for $\lambda_s u^2 > \lambda_h v^2$ and 
``singlet--like'' otherwise. 
The former case corresponds to $\vert \theta\vert >\pi/4$.

One of the mass parameters, say $m_h^2$, can be fixed by requiring
the correct electroweak symmetry breaking,
  $v=246$ GeV.
Then the constraints (\ref{ewbreaking}) specify the allowed
range of 
\begin{equation}
r= {m_s^2 \over m_h^2} \;.
\end{equation}
The required  local minimum exists in the following cases:
\begin{eqnarray}
{\underline {\lambda_{hs} < 0}} &&       \nonumber\\
 && m_h^2 <0~,~m_s^2<0 ~~:~~ 0<r<\infty   ~, \nonumber\\
 && m_h^2 <0~,~m_s^2>0 ~~:~~ \vert r \vert < 
{  \vert   \lambda_{hs}   \vert \over 2 \lambda_h    }~, \nonumber\\
 && m_h^2 >0~,~m_s^2<0 ~~:~~ \vert r \vert > 
{ 2\lambda_s \over  \vert   \lambda_{hs}   \vert    } ~,\nonumber\\
{\underline {\lambda_{hs} > 0}} &&  \nonumber\\
&& m_h^2 <0~,~m_s^2<0 ~~:~~ 
{    \lambda_{hs}    \over 2 \lambda_h } < r < { 2\lambda_s \over    \lambda_{hs}      }~. \label{region}
\end{eqnarray}

 \begin{figure}[ht]
\begin{center}
\epsfig{figure=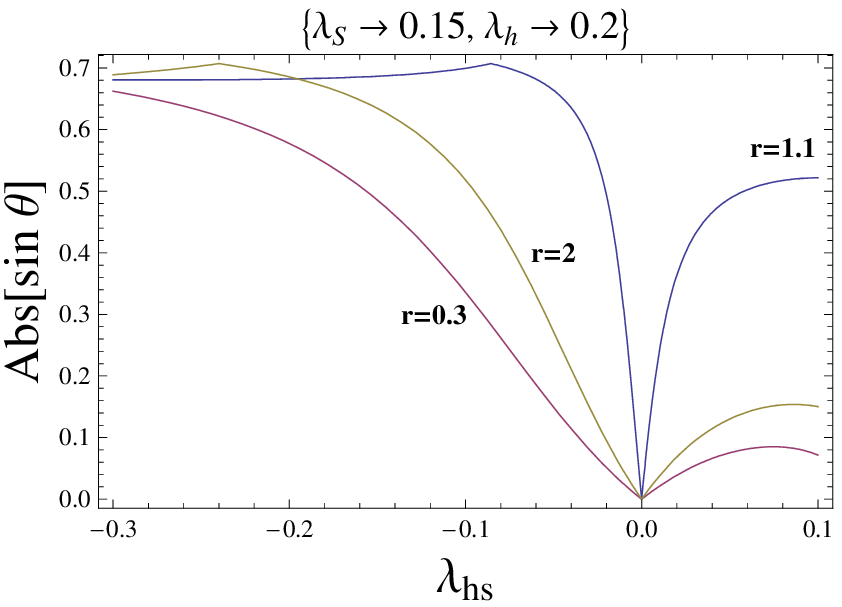,height=5.6cm,width=7.6cm,angle=0}
\hspace{0.3cm} \epsfig{figure=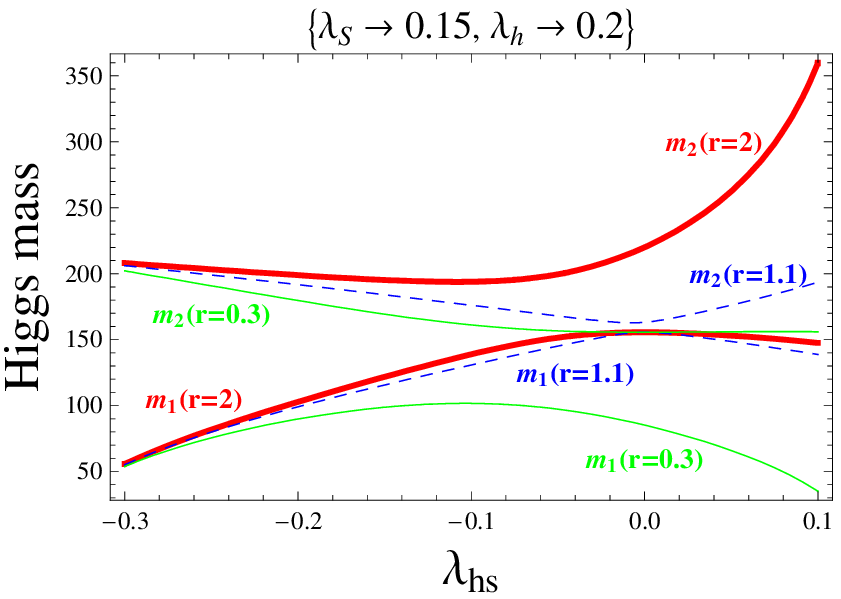,height=5.6cm,width=7.6cm,angle=0}
\end{center}
\medskip
\caption{ 
$\vert \sin\theta\vert$  and the Higgs masses as  functions of $\lambda_{hs}$ 
and $r$ for $m_h^2 <0, m_s^2 <0$. (Here we redefine $\theta$  to be in 
the range $\vert\theta\vert < \pi/4$).
  The parameter range is
consistent with the mixed Higgs--singlet inflaton at $x\sim 1$.  }
\label{fig2}
\end{figure}

We see that at negative $\lambda_{hs}$ there is  more parameter 
space available. In fact, negative values of $\lambda_{hs}$ are 
preferred by the mixed Higgs-singlet inflaton (Fig.~\ref{fig1}),
especially away from the point $x=1$. Indeed, the  relations
among the couplings ensuring $\langle h \rangle \not= 0, \langle s \rangle \not= 0$ 
 at high and low energies are similar up to $\xi_i \leftrightarrow - m_i^2$.   
Representative values of 
the mixing angle consistent with the Higgs-singlet inflaton 
are displayed in Fig.~\ref{fig2}.\footnote{
Eq.~(\ref{tan})  defines $\theta$ up to  $\pi/2$, so in   Fig.~\ref{fig2}
we take  $\vert \theta\vert <\pi/4$. The small kinks in $\sin\theta$ at 
$\lambda_{hs}=-0.24$ and $\lambda_{hs}=-0.08$ correspond to $\vert \tan 2\theta\vert
\rightarrow \infty$, which signals the change in the nature of the lighter 
mass eigenstate. 
In the rest of the paper, we take  $-\sin\theta$ to be  the $h$-component of 
the $H_1$--state.}

Inspection of Eq.~(\ref{eigenvalues}) shows that the lighter 
eigenvalue reaches its upper bound at $\lambda_{hs}=0$. 
In this case, the mixing angle is zero and 
\begin{equation} 
m_1^2=2 \lambda_h v^2, \label{max}
\end{equation}
as in the Standard Model. According to Fig.~\ref{fig1}, this is about
175 GeV. The lower bound on the heavier eigenvalue is also
given by Eq.~(\ref{max}). With the lowest allowed $\lambda_h$,
it     is about 135 GeV.

On the other hand, the heavier eigenvalue can be arbitrarily large.
Indeed, parametrizing
\begin{equation}
u^2 = v^2 ~ {2 \lambda_h r -\lambda_{hs} \over 2 \lambda_s - \lambda_{hs }r }\;,
\end{equation}
we see that $u\rightarrow \infty $ as $r \rightarrow 2\lambda_s /\lambda_{hs}$,
corresponding to the boundary of the region allowed by (\ref{region}).
In this case, $m_2^2 \simeq 2 \lambda_s u^2 \rightarrow \infty$
and the mixing angle approaches zero. 
In terms of the input mass parameters, this corresponds to
$ \vert m^2_{h,s} \vert  \rightarrow  \infty $.
The singlet state can also be arbitrarily light:
in the limit $r \rightarrow \lambda_{hs}/(2\lambda_h) $, $u$ vanishes
and the light eigenstate becomes massless.

\subsubsection{LEP and electroweak  constraints}

LEP has set stringent limits on the Higgs mass and couplings.
For our purposes, the relevant constraint is given in Fig.~10
of \cite{Barate:2003sz}, which sets a bound on 
\begin{equation} 
\zeta^2 \equiv \left(  {g_{HZZ} \over g_{HZZ}^{\rm SM} }  \right)^2 = \vert O_{2i}\vert^2 
\end{equation}
depending on the mass $m_i$.
For a state with an ${\cal O}(1)$  component of $h$, the bound is
\begin{equation}
m > 114~{\rm GeV} \;,
\end{equation}
while for a state with a small admixture of $h$ the bound relaxes
and can be read off from Fig.~10 of \cite{Barate:2003sz}.
For example, with $ \vert O_{2i}\vert^2 \sim 10^{-2} $, the mass
can be as low as 20 GeV. In our case, the bound applies to the lighter
state only since the mass of the 
 heavier state is greater than $\sqrt{2\lambda_h}v > 114$ GeV.  
We therefore require that if $m_1 < 114$ GeV, then
\begin{equation}
\sin^2\theta < \zeta^2 (m_1) \;. 
\end{equation}
For our purposes,  at $\zeta^2<0.5$ it suffices to use an approximation
$\log_{10} \zeta^2 (m) \simeq m/60 - 2.3 $ for $m$ measured
in GeV, which describes the data within a 95\% probability band.

Both mass eigenstates contribute to electroweak observables at a loop level.
For example, the correction to the $\rho$--parameter is \cite{Cerdeno:2006ha}
\begin{equation}
\Delta \rho^H = {3 G_F \over 8 \sqrt{2} \pi^2} \sum_i O_{2i}^2
\biggl(  m_W^2 \ln {m_i^2 \over m_W^2}  -   m_Z^2 \ln {m_i^2 \over m_W^2 }
\biggr) \;.
\end{equation}
This is very similar to the SM Higgs contribution and therefore one can easily
translate the indirect Higgs mass bounds into a bound on 
$\sum_i O_{2i}^2 \ln m_i^2$  \cite{Cerdeno:2006ha,Wells:2008xg}. 
As the benchmark numbers we use the results of  \cite{Erler:2010wa},
$m_H < 148$ GeV (197 GeV) at 95\% (99.5\%) CL. These bounds also incorporate  
results of the LEP and Tevatron direct searches, although purely indirect
constraints give similar numbers \cite{Nakamura:2010zzi}.  
Keeping in mind that  the other oblique as well as    vertex corrections behave 
similar to the $\rho$--parameter  in the heavy Higgs limit and 
 that 
the sensitivity to the Higgs mass is only logarithmic, 
we will use the combined fit results to impose
\begin{equation} 
\sin^2 \theta ~\ln m_1  + \cos^2 \theta  ~\ln m_2 < \ln 148 ~~(197) 
\end{equation}
at 95\% (99.5\%) CL,
where the masses are measured in GeV.

 \begin{figure}[ht]
\epsfig{figure=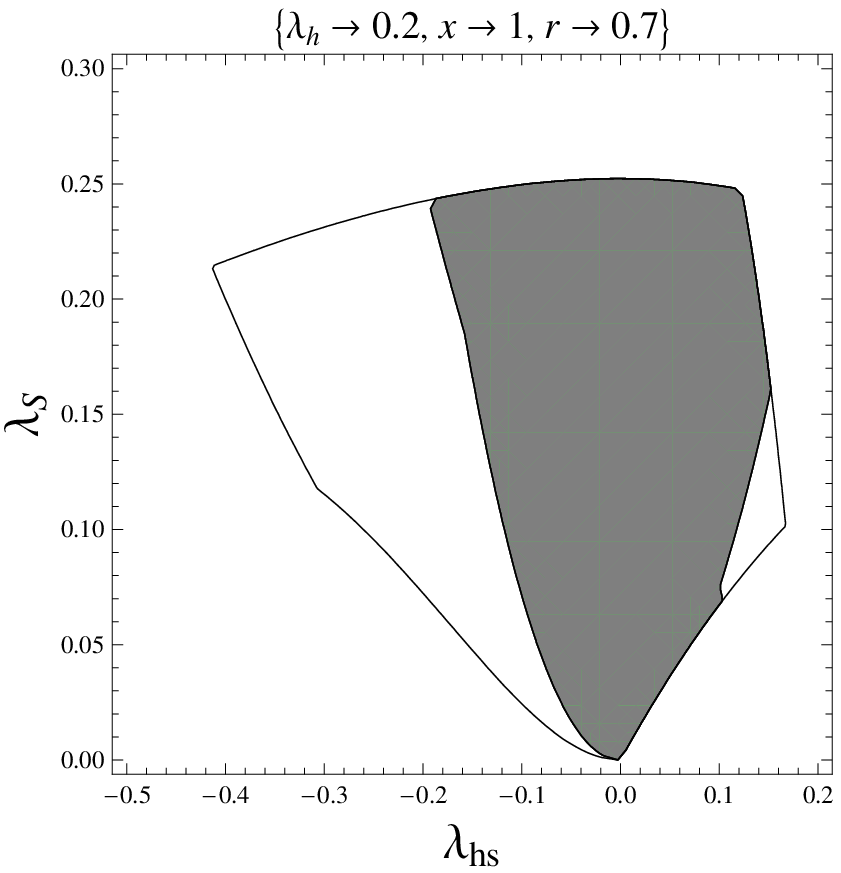,height=5cm,width=5cm,angle=0}
\hspace{0.5cm}\epsfig{figure=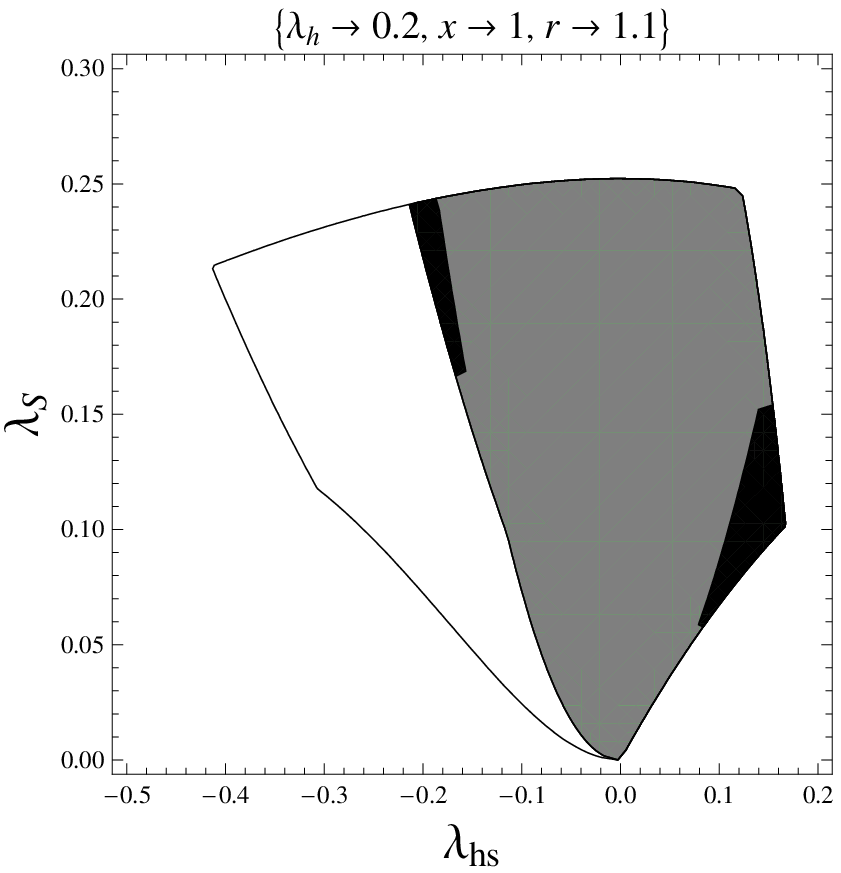,height=5cm,width=5cm,angle=0}
\hspace{0.5cm}\epsfig{figure=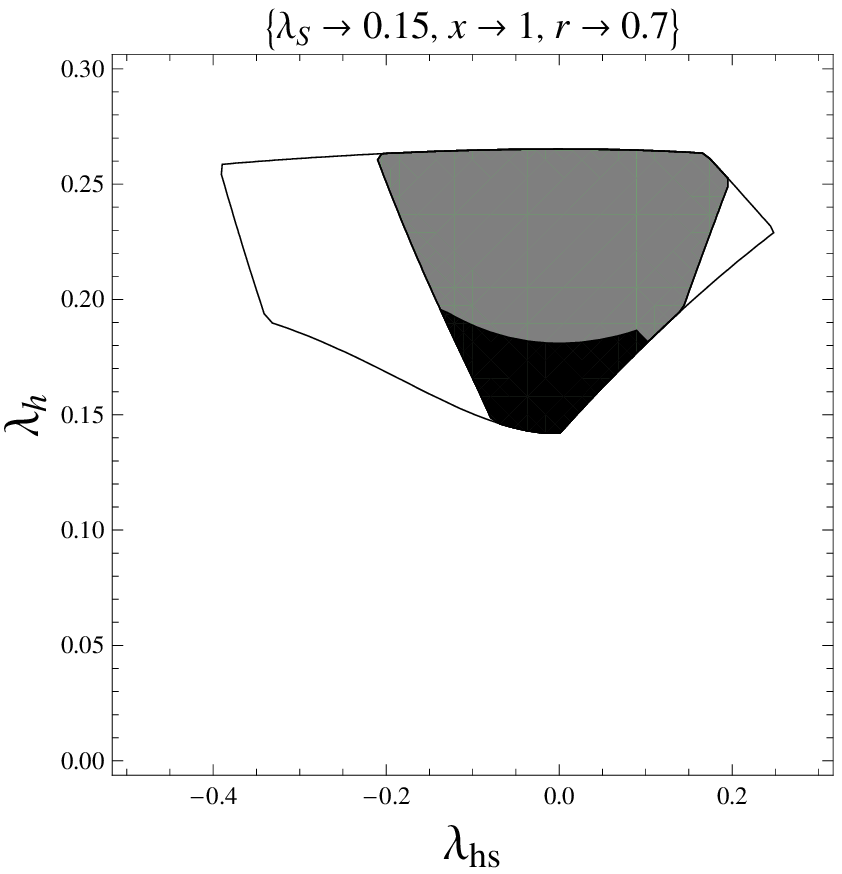,height=5cm,width=5cm,angle=0}
\vspace{0.1cm}\epsfig{figure=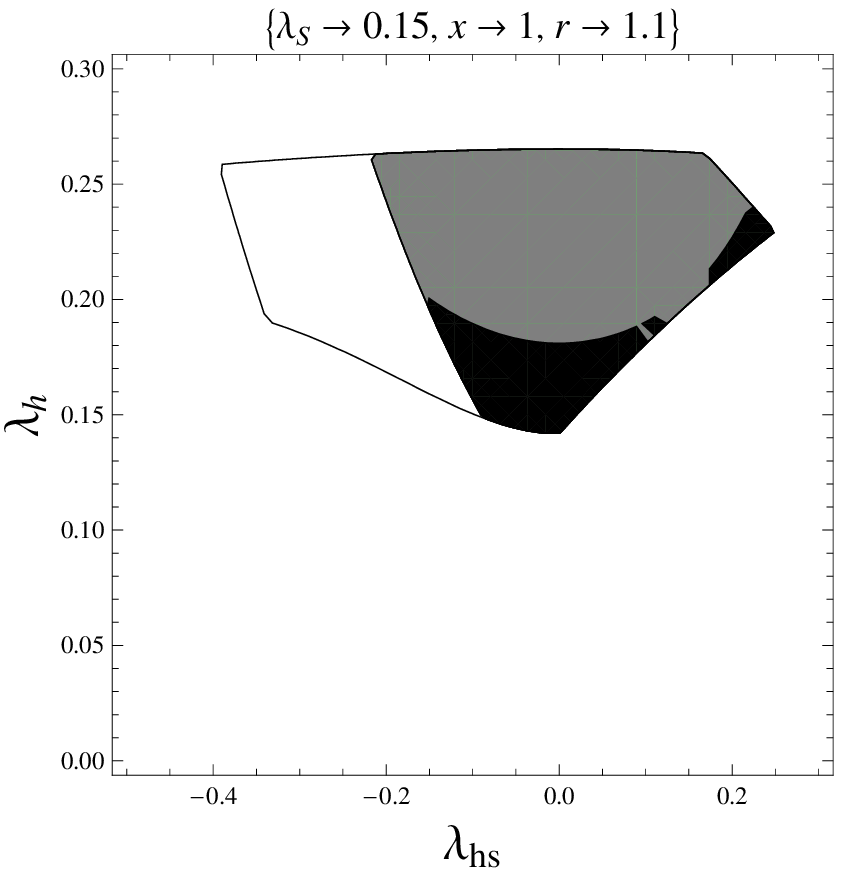,height=5cm,width=5cm,angle=0}
\hspace{0.5cm}\epsfig{figure=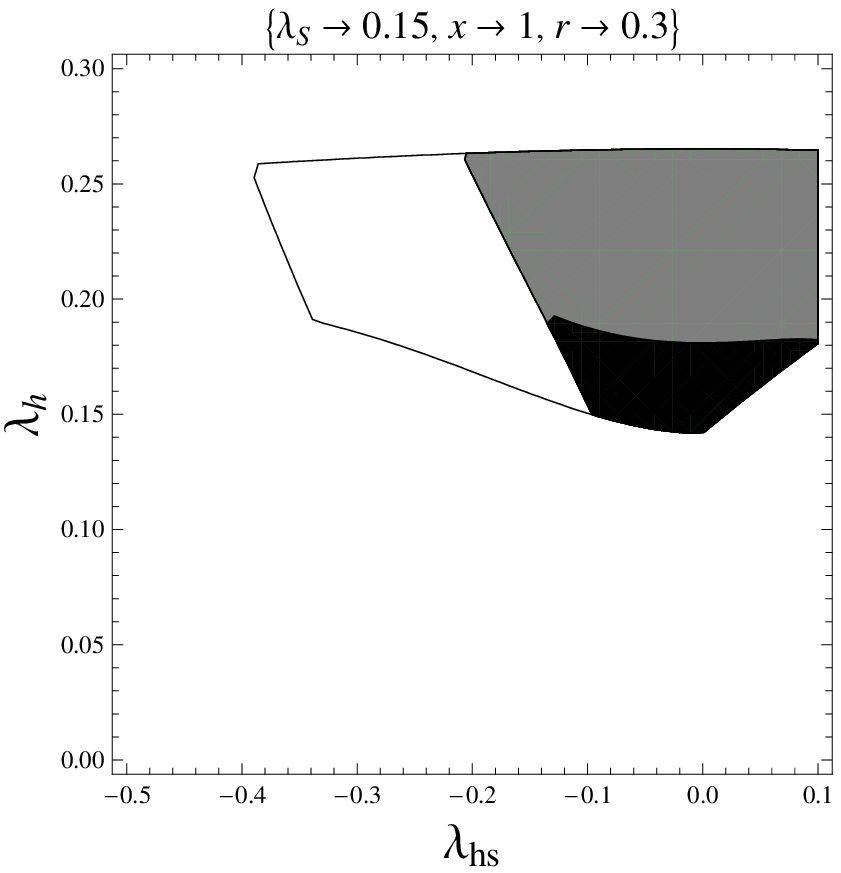,height=5cm,width=5cm,angle=0}
\hspace{0.5cm}\epsfig{figure=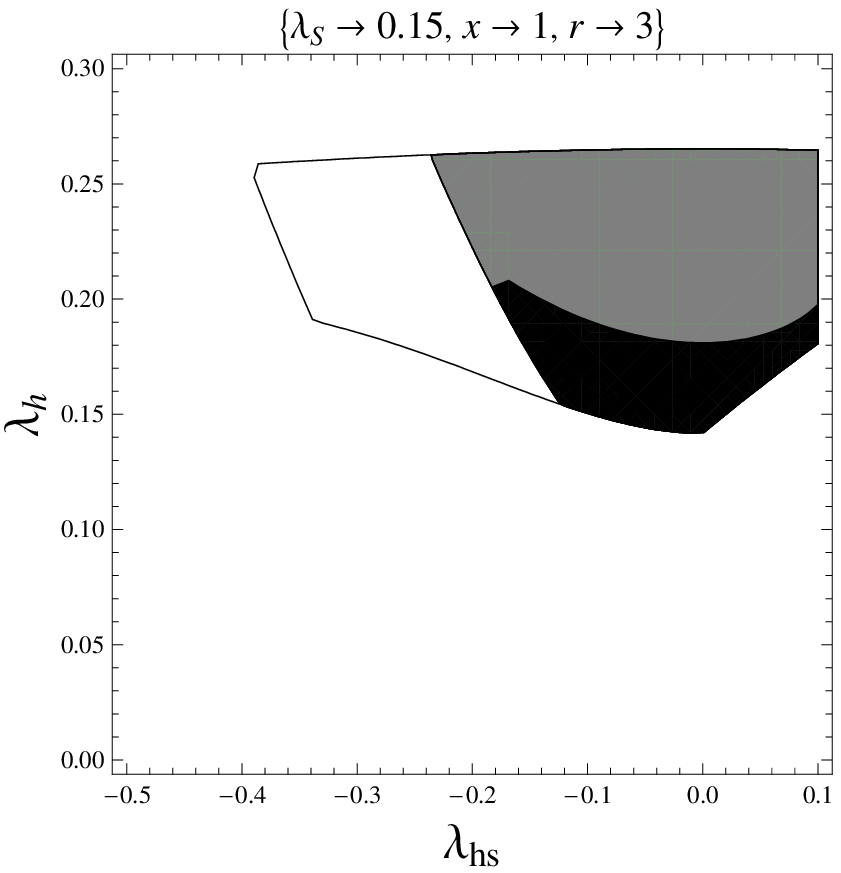,height=5cm,width=5cm,angle=0}
\medskip
\caption{ Parameter space allowed by the LEP and electroweak constraints  
for $m_{s,h}^2 <0$. The region within the contour is allowed by 
the mixed Higgs-singlet inflaton; grey -- allowed by LEP (and automatically
consistent with the 99.5\% CL electroweak constraints); black -- preferred by the 
95\% CL electroweak constraints. $\lambda_i $ are given at the scale $m_t$.   }
\label{fig3}
\end{figure}

The allowed parameter space is presented in Fig.~\ref{fig3}.
The main effect of the LEP constraint is to restrict the size
of $\vert \lambda_{hs} \vert$. The reduction of $\vert \lambda_{hs} \vert$
has a two--fold effect: it decreases the mixing angle and (typically) 
increases the mass of the lighter state (Fig.~\ref{fig2}), 
both of which help satisfy the constraint.
As expected, at larger $\lambda_h$ more  parameter space survives. 
Also, increasing $r$ has a positive effect by 
making  the light state somewhat heavier (Fig.~\ref{fig2}). 

The 99.5\% CL electroweak constraint is satisfied in all the regions
allowed by the LEP bound. However, only a relatively small portion
of parameter space survives the 95\% CL constraint. For instance,
none of the points at $\lambda_s=0.2 , r=0.7$ are allowed. 
Increasing $r$ to 1.1 opens up some parameter space close to the
border of the LEP allowed region. At these points, the nature of the
lighter eigenstate changes compared to the $r=0.7$ case:
it becomes Higgs-like. If the light state is  singlet-like, it is more
difficult to satisfy the EW bound since it is dominated by
the term  $\cos^2 \theta  ~\ln m_2$  with $\cos\theta \sim 1$
and  $m_2 > \sqrt{2 \lambda_h} v$.

In the $\{ \lambda_{hs},\lambda_h  \}$ plane, the preferred region 
is at lower $\lambda_h$, typically $\lambda_h <0.18$.
At $r=0.3$ and $r=3$, the range of $\lambda_{hs}$ must be restricted to
satisfy (\ref{region}). As mentioned above, the composition of the 
lighter state changes with  $r$: it is  typically singlet--like at $r<1$
 and Higgs--like otherwise. Thus, at $r=0.3$ the EW constraint 
is dominated by  $\cos^2 \theta  ~\ln m_2$, while at $r=3$ it is dominated
by $\sin^2 \theta  ~\ln m_1$.

A generalization of the   analysis to $x$ different from 1 is straightforward. 
As clear from Fig.~\ref{fig1}, at $x \gg 1$ or $x\ll 1$, most points at $\lambda_{hs}>0$
get eliminated and negative values of $\lambda_{hs}$ are strongly favored.

The collider signature of the $\langle s \rangle \not= 0$ scenario is a 
universal  suppression of production of the  Higgs--like states,
\begin{equation} 
\sigma (H_i) = \sigma (h)~ \vert O_{2i}\vert^2 \;.
\end{equation}
It is also possible that the decay $H_2 \rightarrow H_1 H_1$ will play a role
 \cite{Bowen:2007ia}. It is kinematically allowed when $\lambda_{hs}$ is considerable 
(see Fig.~\ref{fig2}). Negative  $\lambda_{hs}$ are then largely ruled out by LEP,
while positive $\lambda_{hs}$ are usually consistent with LEP, especially
when $r$ is small or large. For example, at $r=0.3, \lambda_{hs}=0.05$, 
the point 
$m_1=57$ GeV, $m_2=144$ GeV and $\sin\theta=0.07$ is allowed by all the
constraints. 
When $H_2$ is Higgs-like, for $m_2>135$ GeV it will decay 
predominantly into gauge bosons and $H_1$ pairs. The branching
ratio for $H_2 \rightarrow H_1 H_1$ scales like $\lambda_{hs}^2 v^4/m_2^4$
 \cite{Bowen:2007ia,Djouadi:2005gi}, 
which is significant for $\lambda_{hs} > 10^{-1}$ and a light $H_2$.
These values are however disfavored by LEP, so the mode  $H_2 \rightarrow H_1 H_1$ is only competitive below or close to the WW threshold.
In this case, the final state contains 4 b--quarks with relatively low 
 (pairwise)  invariant mass.  
On the other hand, if $H_2$ is singlet--like, its production
cross section is too small and the effect of $H_2 \rightarrow H_1 H_1$
is unimportant.

\subsection{ $\langle h \rangle \not= 0, \langle s \rangle = 0$ 
\label{s0}}

In this case,
\begin{equation}
v^2= - {m_h^2 \over \lambda_h} \;.
\end{equation}
It is a local minimum if
\begin{eqnarray}
&& m_h^2 <0 \;, \nonumber\\
&&  \lambda_{hs} m_h^2 - 2 \lambda_h m_s^2 < 0 \;. 
\end{eqnarray}
There is no mixing between the Higgs and the singlet,
and the mass squared values are
\begin{equation}
 m_1^2= 2 \lambda_h v^2~~,~~  m_2^2= {1\over 2} \lambda_{hs} v^2 + m_s^2 \;.
\label{massesS0}
\end{equation}

The allowed range of $r$ is 
\begin{eqnarray}
{\underline {\lambda_{hs} < 0}} &&       \nonumber\\
  && m_s^2>0 ~~:~~ \vert r \vert > 
{  \vert   \lambda_{hs}   \vert \over 2 \lambda_h    }~, \nonumber\\
 {\underline {\lambda_{hs} > 0}} &&  \nonumber\\
&& m_s^2<0 ~~:~~  r <
{     \lambda_{hs}    \over 2 \lambda_h    } ~, \nonumber\\
&& m_s^2>0 ~~:~~ 0< \vert r\vert < \infty ~. \label{constraints4}
\end{eqnarray}

The analysis of phenomenological constraints is straightforward. Since 
$\lambda_h >0.14$, the Higgs LEP bound is satisfied automatically.
The electroweak precision data favor $\lambda_h <0.18$ (0.32) at
95\% (99.5\%) CL, as in the Standard Model. The allowed parameter space
can then be easily read off from Fig.~\ref{fig1}.

A collider signature of the presence of the singlet ``hidden sector''
would be an invisible decay $h \rightarrow ss$, which for $m_1>
2 m_2$ would typically have a significant branching ratio.    
Note that since $\langle s \rangle = 0$,
the symmetry $s \rightarrow -s $ is not broken spontaneously and the singlet
must be pair--produced. It is relatively easy, especially at small $r$, 
to satisfy the kinematic 
constraint $m_1>2 m_2$: it requires
$ \lambda_{hs}< (1+2r) \lambda_h$ (see Eq.~(\ref{massesS0})).
The corresponding decay width is \cite{Bowen:2007ia}
\begin{equation}
\Gamma (h \rightarrow ss)= { \lambda_{hs}^2 v^2 \over 32 \pi m_1 } \sqrt{1- {4 m_2^2 
\over m_1^2}} \;.
\end{equation}
For $\lambda_{hs}\gg 10^{-2}$, this would be the  dominant decay mode until the 
channel $h \rightarrow WW$ opens up. Above the $WW$ threshold, 
its branching ratio  drops to
${\cal O}( \lambda_{hs}^2 v^4/m_1^4 )  $, 
which  can still be significant for $\lambda_{hs} > 10^{-1}$.
Note that for the $\langle s \rangle =0$ case, larger values of 
$\vert \lambda_{hs}\vert$, up to 0.4, are allowed.

\subsection{LHC prospects}

The Higgs profiling \cite{Englert:2011yb}   at the LHC depends 
crucially on whether or not 
the SM Higgs field mixes with the singlet. If it does, there are two 
states whose masses can be determined by the resonance peak 
measurements. The mixing angle can then be determined by the production
cross section of these states in a particular production mode.
These observables allow us to disentangle 3 quantities 
(up to a sign ambiguity):
\begin{equation} 
m_1 , m_2, \sigma_{\rm prod}  \Rightarrow \lambda_h v^2,
\lambda_s u^2 , \lambda_{hs} u v 
\end{equation}
Furthermore, for a sufficiently heavy $H_2$, measurements of
the cascade
decay $H_2 \rightarrow H_1 H_1$ would determine one more combination of
these quantities such that $u$ can be derived \cite{Englert:2011yb}. 
Since $v$ is known from $M_W$, the couplings 
  $\lambda_h, \lambda_s $ and $\lambda_{hs}$ would then be fixed
(the latter up to the sign). In this case,  the Lagrangian parameters 
are  (almost fully)   reconstructed. Ref.~\cite{Englert:2011yb} provides 
an example  of a point \{$m_1,m_2,\cos^2\theta$\}=\{115 GeV,400 GeV,0.25\},
which can be reconstructed with integrated luminosity 300 fb${}^{-1}$.
The result is  $\lambda_h= 1.04 \pm 0.18$, $u=55.03 \pm 27.35$ GeV,
$\lambda_s = 7.61 \pm 3.51$ and $\lambda_{hs}= 4.52 \pm 2.23$.
This example shows that at least in some regions of parameter space, 
where cascade decays are available,
one can determine  the low energy Lagrangian. 
The precision of this reconstruction grows with integrated luminosity.
Given the low energy parameters, one can evolve them to high energies and
verify whether various inflationary constraints are satisfied. 
In the example above, the perturbativity constraint is violated.
Therefore if such parameter values are indeed found, 
this would falsify the model.

In the case of small or no mixing, the situation is much more challenging,
although the latter option is very interesting as it provides us
with a viable dark matter candidate.
For a heavy singlet, the only possible signature would be missing energy.
If $h \rightarrow ss$ is kinematically allowed, the measurement of the 
Higgs invisible  width would determine $\lambda_{hs}$, up to a kinematical  
factor. For Higgs masses  above 150 GeV, the invisible decay 
into dark matter 
has a small branching fraction and therefore the LHC Higgs exclusion
limits apply (see e.g. \cite{Low:2011kp}). 
For lower Higgs masses, the invisible
decay is efficient and  $\lambda_{hs}$ can be determined using the methods
of \cite{Englert:2011yb}.
However,
the self--interaction coupling $\lambda_s$ is unlikely to be measured
at the LHC, so the Lagrangian cannot be fully reconstructed in this case.

\section{Comparison with the pure singlet or Higgs inflation}

It is instructive to compare the above scenario to the pure singlet or Higgs
inflation. According to Eq.~(\ref{taumin}),   the singlet inflation
($\tau =0 $) requires at high energies
\begin{equation}
 2 \lambda_s \xi_h - \lambda_{hs} \xi_s < 0 \;, \label{singletinfl}
\end{equation}
in which case the ``vacuum'' energy is $\lambda_s/(4 \xi_s^2)$.
This immediately implies 
\begin{equation}
\lambda_{hs} >0 \;.
\end{equation}
The combination $  2 \lambda_h \xi_s - \lambda_{hs} \xi_h    $ can be 
either positive or negative, depending on whether there exists another
local minimum at  $\tau =\infty $. 
We thus leave 
$  2 \lambda_h \xi_s - \lambda_{hs} \xi_h    $   unconstrained.
We further impose the perturbativity and EW vacuum stability
bounds   (\ref{constraints2}).\footnote{Note that singlet inflation
is impossible for negative $\lambda_{hs}$ even if $\xi_s \gg \xi_h$.
In this case, the point $ h=0 , s\rightarrow \infty$ is unstable
and $h$  rolls to infinity. Similarly, Higgs inflation is impossible
for $\lambda_{hs}<0$. For positive $\lambda_{hs}$, our numerical results
are in qualitative agreement with those of \cite{Clark:2009dc,singletinflation}. } 

\begin{figure}[ht]
\epsfig{figure=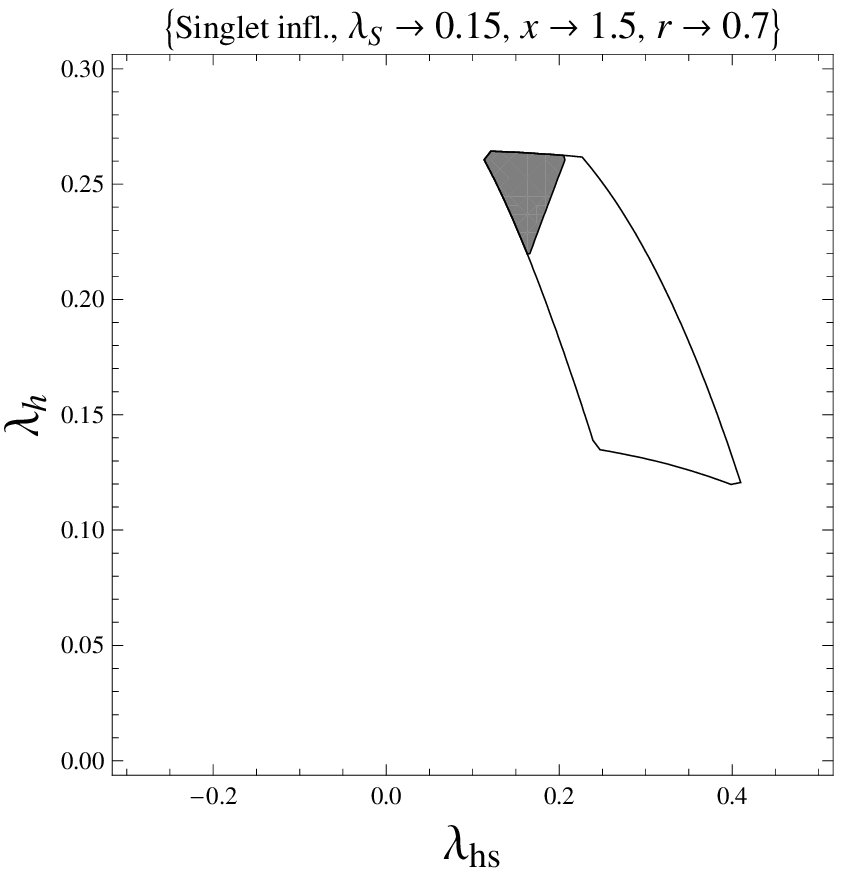,height=5cm,width=5cm,angle=0}
\hspace{0.5cm}\epsfig{figure=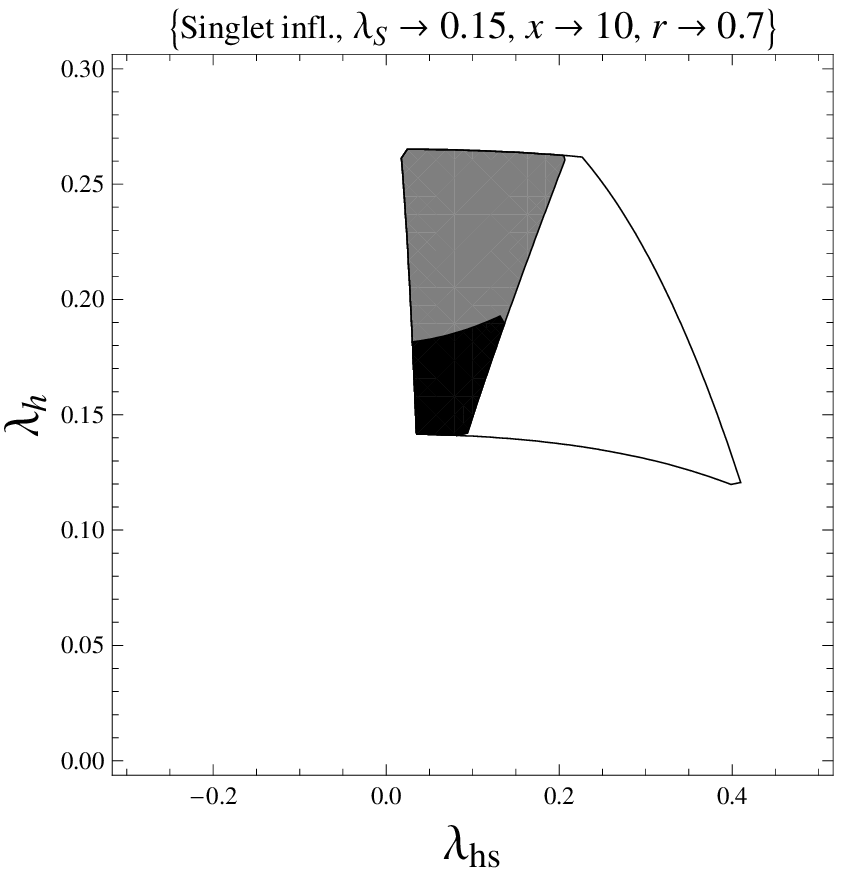,height=5cm,width=5cm,angle=0}
\hspace{0.5cm}\epsfig{figure=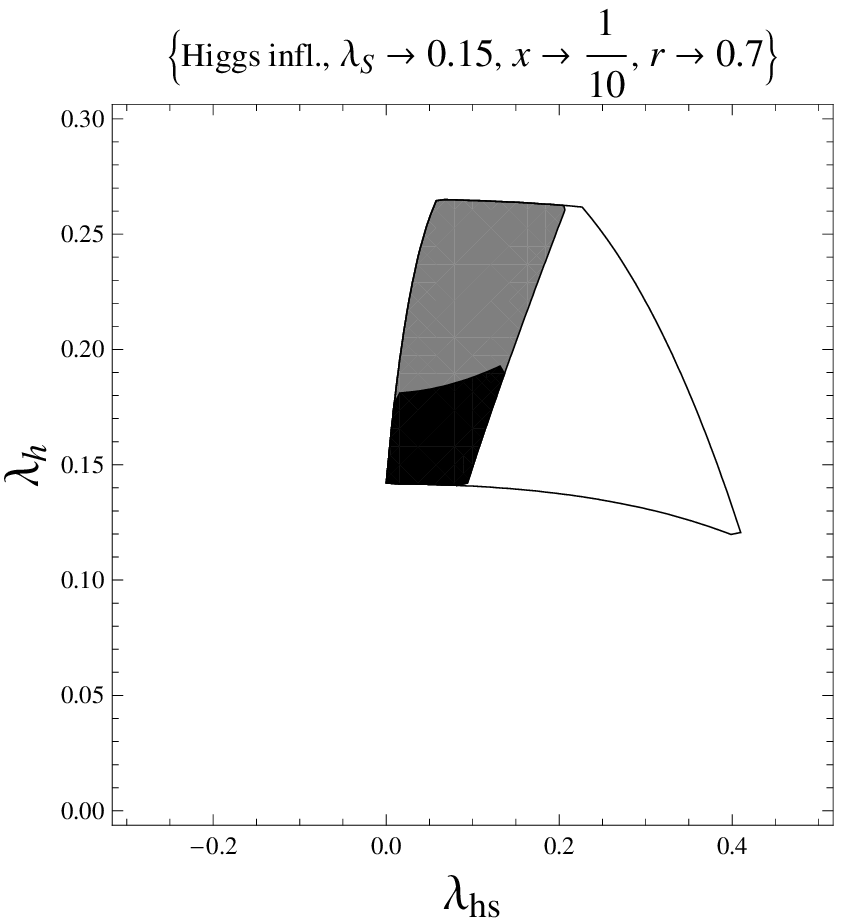,height=5cm,width=5cm,angle=0}
\medskip
\caption{ Constraints on pure singlet and Higgs inflation.
The region within the contour is consistent with singlet (left, center)
and Higgs (right) inflation; grey -- allowed by  $\langle s \rangle \not= 0  $
and LEP; black -- favored by  the 
95\% CL electroweak constraints. Here $m_{s,h}^2 <0$ and $\lambda_i$ are 
given at the scale $m_t$.   }
\label{fig4}
\end{figure}

The phenomenological constraints depend crucially whether or not the singlet
develops a VEV at low energies. For the case $\langle s \rangle \not= 0$,
representative examples are presented in Fig.~\ref{fig4}. The existence of the local
minimum requires $  4 \lambda_s \lambda_h -   \lambda_{hs}^2 >0  $ at low
energies, which together with the LEP Higgs bound eliminates almost all of 
the  parameter space at $x=\xi_s/\xi_h \sim 1$.   For $\lambda_s =0.15$ and  
$x =1.5$, Eq.~(\ref{singletinfl})
requires  $\lambda_{hs}>0.2$ at $\mu_U$. Due to the positive RG contribution
from $\lambda_h$, this bound is easier to satisfy at larger $\lambda_h$,
hence the slanted boundary on the left. Perturbativity and stability further
cut off large values of $\lambda_h, \lambda_{hs}$ and 
small values of $\lambda_h$.

Considerable parameter space is only  available   at $x  \gg 1$,
in which case  (\ref{singletinfl}) amounts to positivity of $\lambda_{hs}$ 
and the allowed region is mainly constrained by the perturbativity and stability 
considerations. 
Since $\lambda_{hs}$ contributes positively to the running of $\lambda_h$,
smaller values of the latter are allowed by $\lambda_h (\mu_U) >0$
at large $\lambda_{hs}$. On the other hand, $\lambda_{hs}$ beyond 0.2
is ruled out by the LEP bound. The 99.5\% CL electroweak precision
constraint is satisfied in the entire region, while the 95\% CL limit  
prefers $\lambda_h$ at the lower end.

If the singlet has a zero VEV, there is no Higgs--singlet mixing and    
the low energy constraints relax. The analysis is very similar to that   
of Sec.~\ref{s0} and phenomenology restricts the values of $\lambda_h$
within the inflation--allowed contours of  Fig.~\ref{fig4}.  For example,
the EW preferred region is $\lambda_h < 0.18$.

Finally, Higgs inflation ($\tau = \infty$) requires
\begin{equation}
 2 \lambda_h \xi_s - \lambda_{hs} \xi_h <0\;
\end{equation}
at high energies,
which again implies $\lambda_{hs} >0$. Significant parameter space
exists only at $x\ll 1$ and the above considerations largely apply,
up to $h \leftrightarrow s$. An example is shown in the right panel of
Fig.~\ref{fig4}.

The main difference between the ``mixed'' and ``pure'' inflaton scenarios lies
in the sign of $\lambda_{hs}$: the former allows for both signs, while the latter
requires a positive   $\lambda_{hs}$.
Note that  $\lambda_{hs} >0$ typically leads to $\langle s \rangle =0$
for a wide range of the parameters,
while $\lambda_{hs} <0$ prefers $\langle s \rangle \not= 0$ 
(see Eqs.~(\ref{region}),(\ref{constraints4})).
Thus the ``pure'' inflation would favor no singlet--Higgs mixing at
low energies and the only collider signature of the singlet would be
an invisible decay $h \rightarrow s s$, if kinematically allowed.
In the mixed inflaton case, $\langle s \rangle =0$ and 
$\langle s \rangle \not= 0$ are almost equally likely. One therefore
often expects Higgs--singlet mixing at low energies which would 
manifest itself in the existence of 2 Higgs--like states with universally
suppressed couplings to the SM fields.

We also observe that, at $\lambda_{hs} >0$, there is an overlap in the allowed parameter space for
 the mixed and pure inflaton (at different $x$), so the collider
data alone may not be sufficient to discriminate among the different
scenarios.\footnote{Presently  it also  seems 
challenging to determine  the sign of $\lambda_{hs}$ at the LHC. One
is likely to need a linear collider to measure scalar self--interactions. }

\section{Unitarity issues} \label{unitaritysection}

The most problematic aspect of Higgs inflation and alike has to do
with unitarity. In the presence of large non-minimal couplings to gravity,
unitarity violation appears around the inflation (Hubble)  scale  
$M_{\rm Pl}/\xi$  \cite{unitarity1,unitarity2}. 
This signals that the theory as it stands is incomplete and 
should be supplemented by additional fields \cite{giudicelee}  
or operators \cite{Lerner:2010mq}   at high energies.

To see how unitarity violation comes about, consider our setup at field values
 $|h|\ll 1/\xi_h$ and $|s|\ll 1/\xi_s$ in Planck units. 
With $\Omega^2$ given in Eq.~(\ref{Omega}), the kinetic terms are 
\begin{equation}
{\cal L}_{\rm kin} = 
{3\over 4} \biggl(  \partial_\mu \log (1+\xi_h h^2 +  \xi_s s^2)    \biggr)^2
+{1\over 2} {1\over 1+ \xi_h h^2 +  \xi_s s^2} \biggl(  (\partial_\mu h)^2 +
(\partial_\mu s)^2   \biggr) \;.
\end{equation}
To leading order in $h\xi_h$ and $s\xi_s$, the mixing between $h$ and $s$
is negligible and we have
\begin{equation} 
{\cal L}_{\rm kin} \simeq \frac{1}{2}(1+6\xi^2_s s^2)(\partial_\mu s)^2 +
\frac{1}{2}(1+6\xi^2_h h^2)(\partial_\mu h)^2 \;. 
\end{equation}
The canonically normalized variables are therefore
\begin{equation}
\rho=s(1+\xi^2_s s^2),\quad \varphi=h(1+\xi^2_h h^2)\;. \label{fielddef}
\end{equation}
We can now expand the fields in terms of expectation values and fluctuations:
\begin{equation}
\rho=\rho_0 + \bar\rho ~,~ \varphi= \varphi_0 + \bar\varphi\;,
\end{equation}
and, similarly, $s=s_0+{\bar s}$ and $h=h_0+{\bar h}$. The fluctuations of the 
original and the canonically normalized fields are related by
 ${\bar s}\simeq (1-3\xi^2_s s_0^2){\bar\rho}-3\xi^2_s s_0\,{\bar\rho}^2$ and
 ${\bar h}\simeq (1-3\xi^2_h h_0^2){\bar\varphi}-3\xi^2_h h_0\,{\bar\varphi}^2$.

Consider  interactions of the Higgs with the gauge bosons. 
The conformal rescaling brings in terms of order $\xi_h h^2$ and 
$\xi_s s^2$, which are negligible compared to $\xi_h^2 h^2$ and 
$\xi_s^2 s^2$. 
 We thus have
\begin{eqnarray}
{\cal L}_{\rm gauge}&=&\frac{1}{2}   g^2h^2 W^+_\mu W^{\mu-} \\
&=&  \frac{1}{2}   g^2\varphi_0^2   \Big(1+2a\frac{{\bar\varphi}}{\varphi_0  }+
b\frac{{\bar\varphi}^2}{\varphi_0^2}\Big)W^+_\mu W^{\mu-}
\end{eqnarray}
with $a=1-3\xi^2_h  \varphi_0^2$ and $b=1-12\xi^2_h \varphi_0^2$. Here we have
neglected the difference between $\varphi_0$ and $h_0$. We see that the 
Standard Model gauge--Higgs interactions  ($a=b=1$)  have changed due to the 
non--canonical normalization. It means that the Higgs exchange no longer unitarizes the $WW$ scattering and the amplitude grows 
with energy:
 ${\cal A}(WW\rightarrow WW) \sim  E^2  \Delta a/\varphi_0^2 
\sim \xi_h^2 E^2$, where $\Delta a$ is the deviation of $a$ from its SM value. 
Thus unitarity is violated at 
$E \sim 1/\xi_h$.

Furthermore, unitarity is violated by scalar interactions. Rewriting the 
Einstein frame scalar
potential in terms of $\varphi$ and $\rho$, we get 
\begin{equation}
U\simeq \frac{1}{4}\lambda_h\varphi^4(1-4\xi^2_h \varphi^2)+\frac{1}{4}\lambda_s \rho^4(1-4\xi^2_s\rho^2)+\frac{1}{4}\lambda_{hs}\varphi^2\rho^2(1-2\xi^2_h\varphi^2-2\xi^2_s\rho^2) \;.
\end{equation}
The 6--point interactions induce $2\rightarrow 4$ scattering with 
a cross section growing as $ E^2 /\Lambda^4$ with $\Lambda= 1/\xi_{s,h}$,
while the unitary bound is $1/E^2$.
Again, for $E>1/\xi_{s,h}$, unitarity is violated.

\subsection{Example of unitarization}

We see that at the scale $1/\xi_{s,h}$ new physics unitarizing 
scattering amplitudes should show up. It may come in a form of new 
degrees of freedom and/or new operators. One possibility is to
complete the theory into a $\sigma$--model by adding a heavy scalar
$\sigma$ \cite{giudicelee}. The corresponding  Jordan--frame  Lagrangian reads
\begin{eqnarray}
{\cal L}_J/\sqrt{-g_J}&=&-\frac{1}{2}(\xi_\sigma {\sigma}^2+{\tilde \xi}_h h^2+{\tilde \xi}_s s^2) R+\frac{1}{2}(\partial_\mu{ \sigma})^2+\frac{1}{2}(\partial_\mu h)^2+\frac{1}{2}(\partial_\mu s)^2 \nonumber \\
&&-\frac{1}{4}\kappa ({\sigma}^2-{\Lambda}^2-\alpha h^2-\beta s^2)^2- V_J(h,s)\;,
\end{eqnarray}
where $V_J(h,s)$ is the Higgs portal potential and $\Lambda= 
1/ \sqrt{\xi_\sigma}$.
Here  the VEV of $\sigma$ generates the Planck scale  (one may also add a
bare $M^2 R$ term \cite{giudicelee})  and 
we take ${\tilde\xi}_h,{\tilde \xi}_s\ll \xi_\sigma$;  
${ \Lambda}\gg v,u$. 
In the low energy limit, 
the heavy $\sigma$--field  
  can be integrated out by minimizing  
the scalar potential (in the Jordan or Einstein frames), 
\begin{equation}
\sigma^2=\Lambda^2+\alpha h^2 +\beta s^2 \;. 
\end{equation}
The resulting effective action is that of the Higgs portal inflation
with effective couplings to gravity 
$\xi_h={\tilde \xi}_h+\alpha \xi_\sigma \simeq \alpha \xi_\sigma$ and 
$\xi_s={\tilde \xi}_s+\beta \xi_\sigma \simeq \beta \xi_\sigma$.

One can easily verify that in the vacuum (at small $u,v$) the canonically
normalized field in the Einstein frame is 
$\chi= \sqrt{6} \ln (\sigma/ \Lambda)$ with mass of order 
$\sqrt{\kappa}/\xi_\sigma$. Substituting $\sigma=\Lambda \exp(\chi/
\sqrt{6})$ back in the potential, one finds that the non-renormalizable
interactions of $\chi$ are Planck--suppressed. On the other hand, since
$\tilde \xi_{h,s} \sim {\cal O}(1)$, unitarity constraints for  interactions
of $h$ and $s$ are satisfied  up to the Planck scale energies.

Let us now consider the inflationary regime $\sigma \gg \Lambda$.
The kinetic terms in the Einstein frame are given by
\begin{eqnarray}
{\cal L}_{\rm kin}&=&\frac{3}{4}\Big[\partial_\mu \ln(\xi_\sigma \sigma^2+{\tilde \xi}_h h^2+{\tilde\xi}_s s^2)\Big]^2 \nonumber \\
&&+\frac{1}{2(\xi_\sigma \sigma^2+{\tilde \xi}_h h^2+{\tilde\xi}_s s^2)}\cdot
\bigg[{(\partial_\mu\sigma)^2}+(\partial_\mu h)^2+(\partial_\mu s)^2\bigg]\;.
\end{eqnarray}
Defining 
\begin{equation}
\chi=\sqrt{3\over 2} \ln (\xi_\sigma \sigma^2) ~,~
\tau_h = {h\over \sigma} ~,~ 
\tau_s = {s\over \sigma} ~,
\end{equation}
we find to leading order in $1/\xi_\sigma$,
\begin{equation}
{\cal L}_{\rm kin}= \frac{1}{2}(\partial_\mu \chi)^2 +
{1\over 2 \xi_\sigma} (\partial_\mu \tau_h)^2 +
{1\over 2 \xi_\sigma} (\partial_\mu \tau_s)^2 \;, 
\end{equation}
while the mixing terms are further suppressed. 
The Einstein frame scalar potential is 
\begin{equation}
U=(\xi_\sigma \sigma^2+{\tilde \xi}_h h^2+{\tilde\xi}_s s^2)^{-2}
\bigg[\frac{1}{4}\kappa (\sigma^2-\Lambda^2-\alpha h^2-\beta s^2)^2+ V_J(h,s)
\bigg]\;,
\end{equation}
which at large $\sigma$ and $\xi_\sigma$ becomes
\begin{equation}
U\simeq \frac{1}{4\xi_\sigma^2} \bigg[\kappa (1-\alpha \tau^2_h-\beta \tau^2_s)^2+\lambda_h\tau^4_h+\lambda_s \tau^4_s+\lambda_{hs}\tau^2_h \tau^2_s \bigg]\;.
\end{equation}
The extremum at $\tau_{h,s}\not= 0$ (``mixed inflaton'')   is given by
\begin{eqnarray}
\tau^2_h&=&\frac{2\kappa (2\alpha\lambda_s -\beta\lambda_{hs})}{4\lambda_h\lambda_s-\lambda^2_{hs}+4\kappa(\alpha^2\lambda_h+\beta^2\lambda_s-\alpha\beta \lambda_{hs})},  \nonumber  \\
\tau^2_s&=&\frac{2\kappa (2\beta\lambda_h -\alpha\lambda_{hs})}{4\lambda_h\lambda_s-\lambda^2_{hs}+4\kappa(\alpha^2\lambda_h+\beta^2\lambda_s-\alpha\beta \lambda_{hs})}\;.
\end{eqnarray}
It is a local minimum if 
\begin{eqnarray}
&&2\alpha\lambda_s-\beta\lambda_{hs}>0, \nonumber  \\
&&2\beta\lambda_h -\alpha\lambda_{hs}>0, \nonumber  \\
&&4\lambda_h\lambda_s-\lambda^2_{hs}+4\kappa(\alpha^2\lambda_h+\beta^2\lambda_s-\alpha\beta \lambda_{hs})>0\;. \label{inf3}
\end{eqnarray}
The last condition follows from the positivity of the determinant of the Hessian.
The value of the potential at this point determines the energy density 
during inflation with  heavy $\tau_{h,s}$  integrated out.   
The resulting inflaton potential is
\begin{equation}
 U(\chi)= {\lambda_{\rm eff}  \over 4\xi_\sigma^2}~
 \Bigl(    1+ {\rm exp}\left( -{2\chi\over \sqrt{6} } \right)          \Bigr)^{-2}
\end{equation} 
with 
\begin{equation}
\lambda_{\rm eff}= \kappa~ \frac{4\lambda_h\lambda_s-\lambda^2_{hs}}{4\lambda_h\lambda_s-\lambda^2_{hs}+4\kappa(\alpha^2\lambda_s+\beta^2\lambda_s-\alpha\beta \lambda_{hs})} \;.
\end{equation}
The denominator of $\lambda_{\rm eff}$ is positive by the stability
condition  (\ref{inf3}), so positivity of the energy density during inflation
requires $4\lambda_h\lambda_s-\lambda^2_{hs} >0 $. Recalling that
$\xi_h \simeq \alpha \xi_\sigma$ and 
$\xi_s \simeq \beta \xi_\sigma$, this condition together 
with (\ref{inf3}) implies
\begin{eqnarray}
&& 2 \lambda_h \xi_s - \lambda_{hs} \xi_h  >0 \;, \nonumber\\
&& 2 \lambda_s \xi_h - \lambda_{hs}  \xi_s >0 \;, \nonumber\\
&&  4 \lambda_s \lambda_h - \lambda_{hs}^2 >0 \;.
\end{eqnarray}
These are exactly the conditions we imposed in our parameter 
space analysis, Eq.~(\ref{constraints1}).\footnote{ Note also that $\alpha^2\lambda_h+\beta^2\lambda_s-\alpha\beta \lambda_{hs}>0$
follows from $2\alpha\lambda_s-\beta\lambda_{hs}>0 , 2\beta\lambda_h -\alpha\lambda_{hs}>0$ (for positive $\alpha,\beta$).} 
Note also that inflation proceeds at the same 
$\tau = \tau_h / \tau_s$ as in the original model. 

Therefore, unitarized Higgs portal inflation leads to the same constraints
on the couplings 
as the original model does. This is despite the fact that now all
three fields participate in inflation, $\tau_{h,s}= {\cal O}(1)$,
and the theory involves an unknown couplings $\kappa$. The latter
affects the energy density, but not the shape of the potential,
so the predictions for the inflationary parameters $n\simeq 0.97$ and
$r\simeq 0.0033$ hold. 
 
\section{Conclusion}

We have studied an extension   of the Higgs sector with a real 
scalar  in the presence of large couplings  to scalar
 curvature. This system supports inflation at large field values,
with tree level predictions $n\simeq 0.97$ and
$r\simeq 0.0033$. 

The nature of the inflaton depends on the relations
among the couplings. For instance, at negative $\lambda_{hs}$,
the inflaton is a mixture of the Higgs and the singlet, while at
positive $\lambda_{hs}$ it can also be a pure Higgs or a singlet.
These requirements leave an imprint on the low energy phenomenology,
e.g. the ``mixed'' inflation often leads to  mixed Higgs--singlet
mass eigenstates at low energies. The latter would manifest themselves at
the LHC as 2 Higgs--like states with universally suppressed couplings.

We have shown how Higgs portal inflation can be unitarized by adding 
an extra scalar with a sub--Planckian VEV. This extension does not however
affect the constraints on the couplings and the low energy phenomenology
remains the same.

{\bf Acknowledgements.}  HML is supported by a CERN--Korean fellowship.


\begin{thebibliography}{999}


\bibitem{inflation}
%\cite{Guth:1980zm}
%\bibitem{Guth:1980zm}
  A.~H.~Guth,
  %``The Inflationary Universe: A Possible Solution To The Horizon And Flatness
  %Problems,''
  Phys.\ Rev.\  D {\bf 23}, 347 (1981);
  %%CITATION = PHRVA,D23,347;%%
%\cite{Linde:1981mu}
%\bibitem{Linde:1981mu}
  A.~D.~Linde,
  %``A New Inflationary Universe Scenario: A Possible Solution Of The Horizon,
  %Flatness, Homogeneity, Isotropy And Primordial Monopole Problems,''
  Phys.\ Lett.\  B {\bf 108}, 389 (1982);
  %%CITATION = PHLTA,B108,389;%%
%\cite{Albrecht:1982wi}
%\bibitem{Albrecht:1982wi}
  A.~Albrecht and P.~J.~Steinhardt,
  %``Cosmology For Grand Unified Theories With Radiatively Induced Symmetry
  %Breaking,''
  Phys.\ Rev.\ Lett.\  {\bf 48}, 1220 (1982).
  %%CITATION = PRLTA,48,1220;%%


%\cite{Bezrukov:2007ep}
\bibitem{Bezrukov:2007ep}
  F.~L.~Bezrukov and M.~Shaposhnikov,
  %``The Standard Model Higgs boson as the inflaton,''
  Phys.\ Lett.\  B {\bf 659}, 703 (2008);
  %%CITATION = PHLTA,B659,703;%%
%\cite{Bezrukov:2010jz}
%\bibitem{Bezrukov:2010jz}
  F.~Bezrukov, A.~Magnin, M.~Shaposhnikov and S.~Sibiryakov,
  %``Higgs inflation: consistency and generalisations,''
  JHEP {\bf 1101} (2011) 016.
  %%CITATION = JHEPA,1101,016;%%


%\cite{Patt:2006fw}
\bibitem{Patt:2006fw}
  B.~Patt and F.~Wilczek,
  %``Higgs-field portal into hidden sectors,''
  arXiv:hep-ph/0605188.
  %%CITATION = HEP-PH/0605188;%%


\bibitem{unitarity1}
%\cite{Burgess:2009ea}
%\bibitem{Burgess:2009ea}
  C.~P.~Burgess, H.~M.~Lee and M.~Trott,
  %``Power-counting and the Validity of the Classical approximation During
  %Inflation,''
  JHEP {\bf 0909} (2009) 103;
  %%CITATION = JHEPA,0909,103;%%
%\cite{Barbon:2009ya}
%\bibitem{Barbon:2009ya}
  J.~L.~F.~Barbon and J.~R.~Espinosa,
  %``On the Naturalness of Higgs Inflation,''
  Phys.\ Rev.\  D {\bf 79}  (2009) 081302  (2009) 081302;
  %%CITATION = PHRVA,D79,081302;%%
%\cite{Hertzberg:2010dc}
%\bibitem{Hertzberg:2010dc}
  M.~P.~Hertzberg,
  %``On Inflation with Non-minimal Coupling,''
  JHEP {\bf 1011}, 023 (2010).
  %%CITATION = ARchiV:1002.2995;%%


\bibitem{unitarity2}
%\cite{Burgess:2010zq}
%\bibitem{Burgess:2010zq}
  C.~P.~Burgess, H.~M.~Lee and M.~Trott,
  %``Comment on Higgs Inflation and Naturalness,''
  JHEP {\bf 1007} (2010) 007.
  %%CITATION = JHEPA,1007,007;%%



%\cite{Giudice:2010ka}
\bibitem{giudicelee}
  G.~F.~Giudice and H.~M.~Lee,
  %``Unitarizing Higgs Inflation,''
  Phys.\ Lett.\  B {\bf 694} (2011) 294.
  %%CITATION = PHLTA,B694,294;%%
  

\bibitem{Clark:2009dc}
  T.~E.~Clark, B.~Liu, S.~T.~Love and T.~ter Veldhuis,
  %``The Standard Model Higgs Boson-Inflaton and Dark Matter,''
  Phys.\ Rev.\  D {\bf 80} (2009) 075019.
  %%CITATION = PHRVA,D80,075019;%%

  
\bibitem{singletinflation}  
%\cite{Lerner:2009xg}
%\bibitem{Lerner:2009xg}
  R.~N.~Lerner and J.~McDonald,
  %``Gauge singlet scalar as inflaton and thermal relic dark matter,''
  Phys.\ Rev.\  D {\bf 80} (2009) 123507.
 %%CITATION = PHRVA,D80,123507;%%


\bibitem{Okada:2010jd}
  N.~Okada and Q.~Shafi,
  %``WIMP Dark Matter Inflation with Observable Gravity Waves,''
  arXiv:1007.1672 [hep-ph];
  %%CITATION = ARXIV:1007.1672;%%
  %\cite{Lerner:2011ge}
%\bibitem{Lerner:2011ge}
  R.~N.~Lerner and J.~McDonald,
  %``Distinguishing Higgs inflation and its variants,''
  arXiv:1104.2468 [hep-ph].
  %%CITATION = ARXIV:1104.2468;%%
  

%\cite{Salopek:1988qh}
\bibitem{Salopek:1988qh}
  D.~S.~Salopek, J.~R.~Bond and J.~M.~Bardeen,
  %``Designing Density Fluctuation Spectra in Inflation,''
  Phys.\ Rev.\  D {\bf 40}, 1753 (1989).
  %%CITATION = PHRVA,D40,1753;%%


%\cite{Lyth:1998xn}
\bibitem{Lyth:1998xn}
  D.~H.~Lyth and A.~Riotto,
  %``Particle physics models of inflation and the cosmological density
  %perturbation,''
  Phys.\ Rept.\  {\bf 314}, 1 (1999).
  %%CITATION = PRPLC,314,1;%%


%\cite{Gong:2011cd}
\bibitem{Gong:2011cd}
  J.~O.~Gong and H.~M.~Lee,
  %``Large non-Gaussianity in non-minimal inflation,''
  arXiv:1105.0073.
  %%CITATION = ARXIV:1105.0073;%%




%\cite{Bezrukov:2008ej}
\bibitem{Bezrukov:2008ej}
  F.~L.~Bezrukov, A.~Magnin and M.~Shaposhnikov,
  %``Standard Model Higgs boson mass from inflation,''
  Phys.\ Lett.\  B {\bf 675}, 88 (2009).
  %%CITATION = PHLTA,B675,88;%%



%\cite{DeSimone:2008ei}
\bibitem{DeSimone:2008ei}
  A.~De Simone, M.~P.~Hertzberg and F.~Wilczek,
  %``Running Inflation in the Standard Model,''
  Phys.\ Lett.\  B {\bf 678}, 1 (2009).
  %%CITATION = PHLTA,B678,1;%%


%\cite{Barvinsky:2008ia}
\bibitem{Barvinsky:2008ia}
  A.~O.~Barvinsky, A.~Y.~Kamenshchik and A.~A.~Starobinsky,
  %``Inflation scenario via the Standard Model Higgs boson and LHC,''
  JCAP {\bf 0811}, 021 (2008);
  %%CITATION = JCAPA,0811,021;%%
A.~O.~Barvinsky, A.~Y.~Kamenshchik, C.~Kiefer, A.~A.~Starobinsky and C.~Steinwachs,
  %``Asymptotic freedom in inflationary cosmology with a non-minimally coupled
  %Higgs field,''
  JCAP {\bf 0912}, 003 (2009).
  %%CITATION = JCAPA,0912,003;%%


%\cite{Bezrukov:2009db}
\bibitem{Bezrukov:2009db}
  F.~Bezrukov, M.~Shaposhnikov,
  %``Standard Model Higgs boson mass from inflation: Two loop analysis,''
  JHEP {\bf 0907}, 089 (2009).


%\cite{Langenfeld:2010aj}
\bibitem{Langenfeld:2010aj}
  U.~Langenfeld, S.~O.~Moch and P.~Uwer,
  %``Measuring the running Top quark mass,''
  arXiv:1006.0097.
  %%CITATION = ARXIV:1006.0097;%%


%\cite{Barate:2003sz}
\bibitem{Barate:2003sz}
  R.~Barate {\it et al.}  [LEP Working Group for Higgs boson searches and
                  ALEPH Collaboration and  and],
  %``Search for the standard model Higgs boson at LEP,''
  Phys.\ Lett.\  B {\bf 565}, 61 (2003).
  %%CITATION = PHLTA,B565,61;%%



%\cite{Cerdeno:2006ha}
\bibitem{Cerdeno:2006ha}
  D.~G.~Cerdeno, A.~Dedes and T.~E.~J.~Underwood,
  %``The Minimal Phantom Sector of the Standard Model: Higgs Phenomenology and
  %Dirac Leptogenesis,''
  JHEP {\bf 0609}, 067 (2006).
  %%CITATION = JHEPA,0609,067;%%


%\cite{Wells:2008xg}
\bibitem{Wells:2008xg}
  J.~D.~Wells,
  %``How to Find a Hidden World at the Large Hadron Collider,''
  arXiv:0803.1243.
  %%CITATION = ARXIV:0803.1243;%%



%\cite{Erler:2010wa}
\bibitem{Erler:2010wa}
  J.~Erler,
  %``The Mass of the Higgs Boson in the Standard Electroweak Model,''
  Phys.\ Rev.\  D {\bf 81}, 051301 (2010).
  %%CITATION = PHRVA,D81,051301;%%


%\cite{Nakamura:2010zzi}
\bibitem{Nakamura:2010zzi}
  K.~Nakamura {\it et al.}  [Particle Data Group],
  %``Review of particle physics,''
  J.\ Phys.\ G {\bf 37}, 075021 (2010);
  %%CITATION = JPHGB,G37,075021;%%
H.~Flacher, M.~Goebel, J.~Haller, A.~Hocker, K.~Monig and J.~Stelzer,
  %``Gfitter - Revisiting the Global Electroweak Fit of the Standard Model and
  %Beyond,''
  Eur.\ Phys.\ J.\  C {\bf 60}, 543 (2009).
  %%CITATION = EPHJA,C60,543;%%




%\cite{Bowen:2007ia}
\bibitem{Bowen:2007ia}
  M.~Bowen, Y.~Cui and J.~D.~Wells,
  %``Narrow trans-TeV Higgs bosons and H ---> hh decays: Two LHC search paths
  %for a hidden sector Higgs boson,''
  JHEP {\bf 0703}, 036 (2007).
  %%CITATION = JHEPA,0703,036;%%




%\cite{Djouadi:2005gi}
\bibitem{Djouadi:2005gi}
  A.~Djouadi,
  %``The Anatomy of electro-weak symmetry breaking. I: The Higgs boson in the
  %standard model,''
  Phys.\ Rept.\  {\bf 457}, 1 (2008).
  %%CITATION = PRPLC,457,1;%%


%\cite{Englert:2011yb}
\bibitem{Englert:2011yb}
  C.~Englert, T.~Plehn, D.~Zerwas, P.~M.~Zerwas,
  %``Exploring the Higgs portal,''
  Phys.\ Lett.\  {\bf B703}, 298-305 (2011).

%\cite{Low:2011kp}
\bibitem{Low:2011kp}
  I.~Low, P.~Schwaller, G.~Shaughnessy, C.~E.~M.~Wagner,
  %``The dark side of the Higgs boson,''
  [arXiv:1110.4405 [hep-ph]].



%\cite{Lerner:2010mq}
\bibitem{Lerner:2010mq}
  R.~N.~Lerner and J.~McDonald,
  %``A Unitarity-Conserving Higgs Inflation Model,''
  Phys.\ Rev.\  D {\bf 82} (2010) 103525.
  %%CITATION = PHRVA,D82,103525;%%





\end{thebibliography}
\end{document}